\documentclass{article} 

\usepackage[a4paper, margin=1in]{geometry} 

\usepackage{amsmath, amssymb, amsfonts}
\usepackage{graphicx}
\usepackage{xcolor}
\usepackage[hidelinks]{hyperref}
\usepackage[section]{placeins}
\usepackage[section]{placeins} 

\title{On the behavior of the Generalized Alignment Index (GALI) method for dissipative systems}

\author{
H.~T.~Moges\\
\small Nonlinear Dynamics and Chaos Group, Department of Mathematics and Applied Mathematics,\\
\small University of Cape Town, Rondebosch 7701, South Africa\\
\small \texttt{MGSHEN002@myuct.ac.za}
\and
T.~Manos\\
\small ETIS Lab, ENSEA, CNRS, UMR8051, CY Cergy-Paris University, Cergy, France\\
\small \texttt{thanos.manos@cyu.fr}
\and
O.~Racoveanu\\
\small European School of Bucharest, Romania\\
\small \texttt{oviracoveanu@gmail.com}
\and
Ch.~Skokos\\
\small Nonlinear Dynamics and Chaos Group, Department of Mathematics and Applied Mathematics,\\
\small University of Cape Town, Rondebosch 7701, South Africa\\
\small \texttt{haris.skokos@uct.ac.za; haris.skokos@gmail.com}
}

\date{} 

\begin{document}

\maketitle

\markboth{H.~T.~Moges et al.}{GALI method for non-Hamiltonian dissipative systems}

\begin{abstract}
The behavior of the Generalized Alignment Index (GALI) method has been extensively studied and successfully applied for the detection of chaotic motion in conservative Hamiltonian systems, yet its application to non-Hamiltonian dissipative systems remains relatively unexplored. In this work, we fill this gap by investigating the GALI's ability to identify stable fixed points, stable limit cycles, chaotic (strange) and hyperchaotic attractors in dissipative systems generated by both continuous and discrete time dynamics, and compare its performance to the analysis achieved by the computation of the spectrum of Lyapunov exponents. Through a comprehensive study of three classical dissipative models, namely the $3$D Lorenz system, a modified Lorenz $4$D hyperchaotic system, and the $3$D generalized hyperchaotic H{\'e}non map, we examine GALI's behavior,  and possible limitations, in detecting chaotic motion, as well as the presence of different types of attractors occurring in dissipative dynamical systems. We find that the GALI successfully detects chaotic motion, as well as stable fixed points, but it faces difficulties in distinctly discriminating between stable  limit cycles, chaotic attractors, and hyperchaotic motion.

\end{abstract}

\textit{Keywords}: Dissipative systems, Lorenz system, hyperchaotic Lorenz system, hyperchaotic H{\'e}non map, Lyapunov exponents, GALI method, chaos, hyperchaos


\section{Introduction} 
\label{sec:Introduction}

It is not that long ago since the huge step into chaos theory came into effect in 1963 when meteorologist Edward Lorenz, trying to perform numerical simulations of the earth's atmosphere, introduced a three-dimensional (3D) autonomous nonlinear chaotic system \cite{lorenz1963deterministic}. Chaotic motion, caused by  nonlinearity in dissipative systems (i.e., systems with the presence of friction or resistance, etc.), is a significant area of study in dynamical systems theory, impacting various disciplines like physics, chemistry and biology (see e.g., \cite{martyushev2006maximum}), engineering \cite{brogliato2007dissipative}, economics \cite{zhang2006discrete} and communication security \cite{grassi1999system,cuomo1993circuit}. After R\"{o}ssler published his work on the analysis of a hyperchaotic system, that is, a system having two positive Lyapunov exponents (LEs), in 1979 \cite{rossler1979equation}, numerous researchers delved into the study of hyperchaotic systems. Furthermore, in \cite{baier1990maximum}, the authors introduced a discrete hyperchaotic map derived from the well-known H{\'e}non map \cite{henon1976two}. Subsequent to these developments, various well-known chaotic and hyperchaotic systems of both continuous and discrete time have been proposed, as documented in \cite{sprott2010elegant}. A recent review article collection \cite{awrejcewicz2021introduction} offers an overview of the latest advancements in modeling complex systems exhibiting chaotic and hyperchaotic behavior. Hyperchaotic systems offer enhanced randomness and unpredictability compared to chaotic systems in areas of real-world applications such as image encryption (see e.g., \cite{wang2023image}), as the added complexity in parameter space increases security by making it harder for attackers to decipher the encryption method or extract information from encrypted data.

Throughout the years, LEs (see e.g., \cite{ benettin1980lyapunova,benettin1980lyapunovb,Skokos2010}) have served as a valuable tool to characterize the asymptotic behavior of nonlinear dynamics, and are widely employed for identifying chaotic behavior. In addition, several methods including the Fast Lyapunov Indicator (FLI) \cite{froeschle1997fastb, froeschle1997fasta, lega2016theory}, as well as the Smaller (SALI) \cite{skokos2001alignment,skokos2003does,skokos2004detecting} and the Generalized Alignment Index (GALI) \cite{skokos2007geometrical,Skokos2016} methods, have been introduced over time to discern between regular and chaotic motion. The GALI method stands out as a well-established and effective numerical technique for detecting chaos in Hamiltonian systems and area-preserving maps \cite{skokos2007geometrical, skokos2008detecting, bountis2009application, manos2012probing, BCSPV12, Skokos2016}, having significant advantages over the most commonly used chaos detection method, i.e., the computation of the maximum LE (mLE), as it overcomes the slow convergence of the mLE to its limiting value.  Additionally, the GALI indices have been proven powerful in  identifying the dimensionality of tori on which regular motion takes place \cite{skokos2008detecting, Moges2020}.

Up until now, the SALI and the GALI methods have predominantly been applied to the investigation of conservative systems. Nevertheless, there have been some preliminary applications of these methods to non-autonomous dissipative models \cite{huang2012analysis, huang2013circuit, tchakui2020chaotic}, while in \cite{huang2013circuit} both the SALI and the FLI methods were used to identify parameter intervals associated with ordered or chaotic trajectories in a modified Lorenz chaotic system. In addition, the exploration of the transition from regular to chaotic behavior via a model parameter variation, has been conducted for a modified L\"{U} chaotic system having exponential terms in \cite{huang2014numerical}. Furthermore, the chaotic properties of a five-dimensional ($5$D) fractional-order chaotic system were examined using LEs and the SALI in \cite{yan2023analysis}. These studies showed SALI's efficacy as a chaos detection technique for dissipative dynamical systems. Moreover, the GALI method has also been used for the detection of regular and chaotic motion in Hamiltonian systems having a relatively slow time-dependency when one or more system parameters vary with time \cite{manos2013interplay, ManosMachado2014, MachadManoso2016, tchakui2020chaotic, ManSkoPa2021}. 

Our work primarily aims to systematically explore the performance of the GALI method in dissipative dynamical systems of continuous time whose evolution is generated by a set of ordinary differential equations (ODEs), and of discrete time dissipative maps. Our objective is to investigate the behavior  of the GALI indices for various typical cases of trajectories observed in such dissipative systems. To this end, we initially identify the possible different types of motion that can be encountered in dissipative systems by computing the respective spectrum of LEs per case, and then we compute the time evolution of the various GALI indices. We perform our investigations for three classical, well-studied models, each exhibiting distinct dynamical features. We begin by considering the classical 3D Lorenz system, renowned for the presence of strange attractors \cite{lorenz1963deterministic}. Then, we study a modified Lorenz four-dimensional (4D) dissipative chaotic system \cite{yujun2010new} to investigate hyperchaotic motion, i.e., trajectories having  two (or more) positive LEs. Finally, we consider the 3D generalized hyperchaotic H{\'e}non map \cite{awrejcewicz2018quantifying}, whose two control parameters give rise to both chaotic and hyperchaotic attractors. By using these three models, we examine the dynamics of both continuous and discrete time dynamical systems and provide a thorough investigation of the GALI's behavior.

The paper is organized as follows. The LEs and the GALI methods are introduced in Sec.~\ref{sec:Methods}. The three dissipative dynamical systems considered in our investigation are introduced in Sec.~\ref{sec:Models}. The behavior  of the LEs and the GALI method for several trajectories of these models, as well as a comparison between the results obtained by these chaos indicators, are presented  in detail in separate subsections of Sec.~\ref{sec:Results}; namely, the 3D Lorenz system is studied in Sec.~\ref{sec:3DODE}, the 4D continuous hyperchaotic system is analyzed in Sec.~\ref{sec:4DODE}, and the generalized hyperchaotic H{\'e}non map is investigated in Sec.~\ref{sec:3DHenMap}. Finally, the main findings and conclusions of our work are summarized in Sec.~\ref{sec:Summary}.

\section{Chaos detection techniques} 
\label{sec:Methods}

We introduce here the two computational methods  we use in this paper to distinguish between regular and various types of chaotic motion occurring in the three dissipative systems we study.  In order to classify the different types of chaotic motion and attractors encountered in these systems, we utilize the full spectrum of LEs. Hence, we start by briefly providing the definition of the LEs and then we define the GALI method. In order to avoid  repetitions, we define the LEs and the GALI method only for continuous time dissipative systems, and make explicit comments whenever  these definitions need adaptation for  discrete time systems.

We start by considering a trajectory of a  conservative autonomous dynamical system whose evolution in the system's $N$-dimensional ($N$D) phase is governed by a set of ODEs \cite{taylor2005classical}:
\begin{equation}
	\dot{\textbf{x}} = \frac{d\textbf{x}}{dt} = \textbf{f}(\textbf{x}(t)),
	\label{eq:dynamics_cont}
\end{equation}
where the vector $\textbf{x}(t) \in \mathbb{R}^N$ represents the state variables, and $\textbf{f}: \mathbb{R}^N \longrightarrow \mathbb{R}^N$ is a vector field. In \eqref{eq:dynamics_cont}  $\dot{\textbf{x}}$ denotes the time-derivative $\dfrac{d\textbf{x}}{dt}$. Equation \eqref{eq:dynamics_cont} can be understood as describing the evolution of a dynamical system defined by a finite-dimensional state vector $\textbf{x}(t)$ of dimension $N$, which evolves continuously over time $t$.

To define the LEs and the GALI, the concept of variational equations is needed (see for example \cite{Skokos2010} and references therein for more details). These equations represent the linearized version of ODEs governing the time evolution of an infinitesimal perturbation $\textbf{v}(t)$ (typically called a deviation vector) of  a reference trajectory $\textbf{x}(t)$, and have the form:
\begin{equation}
	\dot{\textbf{v}}(t) = \textbf{J}(\textbf{x}(t)) \cdot {\textbf{v}}(t_0),
	\label{eq:Ham_Eqn_variational} 
\end{equation}
where $\textbf{J}(\textbf{x}(t))$ is the Jacobian matrix of $\textbf{f}(\textbf{x}(t))$ and $\textbf{v}(t_0) = (\delta x_1(t_0), \dots, \delta x_N(t_0))$ is the initial deviation vector from a given trajectory with  initial condition (IC) $x(t_0)$ at the starting time $t_0$. The components of $\textbf{v}(t_0)$, denoted as $\delta x_1(t_0), \dots, \delta x_N(t_0)$, represent small perturbations introduced to each one of the state variables of the system. For continuous systems, the evolution of the vector $\textbf{v}(t_0)$ is computed through the simultaneous integration of the sets of ODEs given in \eqref{eq:dynamics_cont} and \eqref{eq:Ham_Eqn_variational}.  The vector $\textbf{v}(t)$ essentially describes how small perturbations from the IC evolve over time along the trajectory of the dynamical system.

We note that the divergence, $\mathbf{\nabla} \textbf{f}$, of the vector field $\textbf{f}$ in \eqref{eq:dynamics_cont}  determines the instantaneous rate of change of the phase space volume along the trajectory $\textbf{x}(t)$, and can be either positive (expanding),  negative (contracting) or zero (conserved). The  average  (over time)  rate of phase space volume change, $\Delta \textbf{f}$, can be used to determine  whether a dynamical system is dissipative or not. In particular, a system is characterized as dissipative, conservative or expanding in phase space volume, if $\Delta \textbf{f} <0$, $\Delta \textbf{f} =0$ or $\Delta \textbf{f} >0$, respectively \cite{taylor2005classical}. We emphasize that  the three dynamical systems considered in our study are dissipative.

\subsection{Lyapunov Exponents} 
\label{sec:LEs}

A $N$D dynamical system has $N$ LEs. Among these, the mLE, $\chi_1$, essentially measures the average rate of convergence and divergence between neighboring trajectories in the dynamical system's phase space, and it is typically computed as \cite{benettin1980lyapunova, 
 benettin1980lyapunovb, Skokos2010}:
\begin{equation} 	
\label{eq:mLEs_chi}
	\chi_{1} = \lim\limits_{t\to \infty} \lambda_{1}(t), 
\end{equation}
where the quantity $\lambda_{1}$ is the so-called  \textit{finite time maximum Lyapunov exponent} (ftmLE)   defined as: 
\begin{equation}
	\lambda_{1}(t) = \frac{1}{t}\ln \frac{\lVert{\textbf{v}(t)}\Vert}{\lVert{\textbf{v}(0)}\Vert}.
	\label{eq:mLEs_lamb}
	\end{equation}
In \eqref{eq:mLEs_lamb} vectors $\textbf{v}(0)$ and $\textbf{v}(t)$ represent the deviation vectors from a given trajectory at times $t=0$ and $t>0$, respectively, while $\Vert{ \cdot }\Vert$ denotes the usual Euclidean norm of a vector. Similarly, the other LEs, $\chi_{2}$, $\chi_{3}$, $\cdots$, $\chi_{N}$  ($\chi_{1} \geq \chi_{2} \geq \chi_{3} \geq \cdots \geq \chi_{N}$), can be computed as the $t \rightarrow \infty$ limits of appropriately defined quantities, $\lambda_{2}(t)$, $\lambda_{3}(t)$, $\cdots$, $\lambda_{N}(t)$, which are called \textit{finite time LEs} (ftLEs) (see \cite{Skokos2010} for more details). In this work, we follow the numerical algorithm proposed by \cite{benettin1980lyapunova, benettin1980lyapunovb} to compute the whole set of LEs of a dynamical system, i.e., the so-called  spectrum of LEs. 

The mLE is used to distinguish between regular ($\chi_{1} = 0$) and chaotic ($\chi_{1} > 0$) motion. We note that in Hamiltonian systems the mLE is strictly positive for ICs leading to chaotic motion, while for regular trajectories  the ftmLE \eqref{eq:mLEs_lamb} tends to zero following a power law decay $\lambda_1(t) \propto t^{-1}$  \cite{Skokos2010}. Furthermore, computing more (or even all) LEs offers additional insights into the underlying dynamics and the statistical properties of a dynamical system.  In cases of hyperchaotic motion, the two (or more) largest LEs are positive. Furthermore, the spectrum of LEs can be utilized to characterize various types of motions, e.g., limit cycles, chaotic strange attractors, and hyperchaotic attractors. More specifically, in the $N$D phase space of a dynamical system we can have various types of trajectories and attractors \cite{cencini2010chaos}:
\begin{enumerate}
	\item \textit{Stable fixed point}: In this case all  LEs are negative. This arrangement is represented as $(-, -, \cdots ,-)$.
	\item \textit{Stable limit cycle}: For a limit cycle, the mLE is zero, while the remaining LEs are all negative. This case is denoted as $(0, -,\cdots, -)$.
	\item \textit{$k$-dimensional stable torus}: Here, the first $k$ LEs are equal to zero, while the remaining ones are negative. We denote this arrangement of LEs as $(k(0,\cdots, 0), -,\cdots,-)$.
	\item \textit{Chaotic strange attractor}: For a strange attractor the mLE is positive,  the second largest LE is zero, while the rest are negative. This arrangement is represented as $(+, 0, -,\cdots,-)$.
	\item \textit{Hyperchaotic attractor}: In this case, there are at least two positive LEs. 
\end{enumerate}

It is worth mentioning that the sum of all LEs ($\sum_{j=1}^{N} \chi_j$) measures the average contraction rate of phase space volumes. In dissipative systems, the phase space volume formed by a set of trajectories undergoes exponential shrinking, hence resulting in a negative sum of LEs.

\subsection{The GALI method}
\label{sec:GALI}

The SALI and GALI methods have been used for over two decades as effective chaos detection techniques for dynamical systems. The GALI of order $k$ (GALI$_k$) is a measure that quantifies the volume of a generalized parallelogram formed by $k$ unit deviation vectors, $\hat{\textbf{v}}_1, \hat{\textbf{v}}_2, \cdots, \hat{\textbf{v}}_k$, and is computed as the norm of the wedge product of these vectors \cite{skokos2007geometrical}:
\begin{equation} \label{Def:GALI}
\mbox{GALI}_k(t) = \lVert \hat{\textbf{v}}_1 \wedge \hat{\textbf{v}}_2 \wedge \hat{\textbf{v}}_3 \wedge \dots \wedge \hat{\textbf{v}}_k \Vert.
\end{equation}
Let us briefly discuss the behavior of the GALI for $N$D Hamiltonian systems.  For regular trajectories any $k \le N$ linearly independent initial deviation vectors used to compute the GALI$_k$ will eventually fall on the $N$D tangent space of the torus on which the motion takes place \cite{skokos2007geometrical}. In this case, the  GALI value remains positive and  practically constant, i.e.:
\begin{equation}
\label{eq:GALI_regular}
\mbox{GALI}_k(t) \propto \text{constant}, \quad \text{if} \quad k \le N.
\end{equation}
If we consider  $k > N$ linearly independent initial deviation vectors  the asymptotic GALI value will be zero,  because the set of  deviation vectors will eventually become linearly dependent, as they will all fall on the $N$D tangent space of the torus. In this case, the GALI$_k$ tends to zero following a well defined power law decay \cite{skokos2007geometrical}. On the other hand, for chaotic trajectories (and unstable periodic orbits)  all deviation vectors will eventually align to the direction defined by the mLE and consequently the value of the GALI$_k$ decays exponentially fast to zero. The rate of this decay depends on the values of the $k$ largest LEs, as detailed in \cite{skokos2007geometrical,manos2012probing}:
\begin{equation}
\label{eq:GALI_chaos}
\mbox{GALI}_k(t) \propto e^{-[(\chi_1-\chi_2)+(\chi_1-\chi_3)+\dots+(\chi_1-\chi_k)]t}.
\end{equation}

\section{Continuous and discrete time dynamical models}
\label{sec:Models}

To explore the behavior of the GALI indices for  various types of attractors in continuous and discrete time dissipative systems, we consider  three simple dynamical models  (two continuous time systems and one discrete time map). These dynamical models allow us to cover different characteristic types of trajectories in our investigation, including stable fixed points, stable limit cycles, chaotic  and hyperchaotic attractors. To the best of our knowledge,  no comprehensive, detailed studies (similar to the  ones performed for LEs in e.g., \cite{qi2005analysis, Zheng2018}) have been conducted to investigate how the GALI indices behave for different types of attractors in dissipative systems. 

In our investigation of the continuous time dissipative systems (Secs.~\ref{sec:3DODE} and \ref{sec:4DODE}) we implement the fourth-order Runge–Kutta integration method to numerically solve the equations of motion, along with their corresponding variational equations. However, when we consider a dissipative discrete map in Sec.~\ref{sec:3DHenMap} we  compute the evolution of the studied trajectory and its associated deviation vectors by iterating both the map itself along with the so-called tangent map, which governs the deviation vector's evolution (see e.g.,  \cite{Skokos2010}). 

\subsection{The 3D Lorenz system}
\label{sec:Models_3D_Lorenz}

We start our analysis by considering  one of the most famous dissipative systems, namely the 3D Lorenz system \cite{lorenz1963deterministic}, which  is a well-known example of a chaotic, nonlinear dynamical system that has been broadly studied in the literature (see e.g., \cite{Lorenz1995, Strogatz2018, Shen_IJBC_2023_review}). The system consists of three coupled ODEs that describe the behavior of a simplified atmospheric convection model:
\begin{eqnarray} 
	\label{eq:3DODE} 
	\dot{x} &=& a(y -x), \nonumber \\
	\dot{y} &=& rx - y - xz,  \\
	\dot{z} &=&  xy - bz, \nonumber
\end{eqnarray}
where $x$, $y$ and $z$ are state variables and $a$, $b$ and $r$ are parameters controlling the trajectories' dynamical evolution. For instance, when $a =10$ and $b=8/3$ system \eqref{eq:3DODE} exhibits chaotic attractors of different shapes depending on the value of $r$. More specifically, this model will help us analyze and compare the behavior of the LEs and the GALIs for stable fixed points, limit cycles, and chaotic motions.\\

\subsection{The 4D Lorenz hyperchaotic system}
\label{sec:Models_4D_Lorenz}

The next model we consider is a modified 4D Lorenz hyperchaotic system \cite{yujun2010new} which  exhibits hyperchaotic motion (i.e., trajectories having two positive LEs). The model originates from the standard 3D Lorenz system \eqref{eq:3DODE}, with the addition of an extra nonlinear term in the set of ODEs, which also include a  feedback control term and a coupling term  (see e.g., \cite{qi2005analysis}). The system is defined by a set of four coupled ODEs as follows:
\begin{eqnarray} 
\label{eq:4DODE} 
\dot{x} &=& a(y -x) + yz, \nonumber \\
\dot{y} &=& cx - y - xz + w,  \\
\dot{z} &=&  xy - bz,  \nonumber \\
\dot{w} &=&  -xz + rw, \nonumber
\end{eqnarray}
where $x$, $y$, $z$ and $w$ represent the state variables of the system, and $a$, $b$, $c$ and $r$ are its control parameters. In \cite{yujun2010new} it was demonstrated that this system can display  hyperchaotic motion by fine-tuning its parameter $r$. For example, by setting $a = 35$, $b = 8/3$ and $c = 55$ model \eqref{eq:4DODE} exhibits a wide range of chaotic behaviors characterized by a positive mLE, along with hyperchaotic motion having  two positive LEs. We note that for these parameter values,  system \eqref{eq:4DODE} is dissipative when  $r < 38.667$ \cite{yujun2010new} and, for instance, $r=1.3$ results in the appearance of  hyperchaotic behavior.

\subsection{The generalized hyperchaotic H{\'e}non map}
\label{sec:Models_Henon_map}

The last model we consider is a generalized version of the classical two-dimensional (2D) H{\'e}non map \cite{henon1976two}, the so-called generalized hyperchaotic H{\'e}non map \cite{baier1990maximum}, which has an additional state variable, $z$, compared to the original model. The 3D H{\'e}non map is described by the following equations: 
 \begin{eqnarray}
	\label{eq:3DHenMap} 
	x' &=& a - y^2  - b z, \nonumber \\
	y' &=& x,  \\
	z' &=& y, \nonumber
\end{eqnarray}
where $x$, $y$ and $z$ are the state variables at discrete time $n$,  $a$ and $b$ are the control parameters of the map and $(x', y', z')$ denotes the evolved state vector after one iteration of the map. Map  \eqref{eq:3DHenMap} exhibits a richer dynamics with respect  to its 2D counterpart, including the presence of hyperchaotic motion for $a=1.6$ and $b=0.01$.

\section{Numerical results}
\label{sec:Results}

In this section, we examine in detail the  behavior of the GALI method for various representative types of trajectories appearing in the three considered dynamical systems described in Sec.~\ref{sec:Models}. To this end, we consider stable fixed points, stable limit cycles, chaotic strange attractors, and hyperchaotic trajectories. Let us also stress that the choice of the models' parameters resulting in these diverse dynamical behaviors and motions, was achieved   with the help of the computation of the related LEs spectra.

\subsection{Numerical investigation of the 3D Lorenz system} 
\label{sec:3DODE}

In what follows, we fix the parameters of the 3D Lorenz system \eqref{eq:3DODE} to $a=10$ and $b=8/3$ and allow the third parameter, $r$,  to vary. For our numerical simulations, we integrate  system \eqref{eq:3DODE} and its respective variational equations obtained by the application of \eqref{eq:Ham_Eqn_variational}, up to $t=10^5$ time units.

\subsubsection{A stable fixed point case}
\label{sec:3DODE_fp}

\begin{figure}\centering
	\includegraphics[width=1\columnwidth, keepaspectratio]{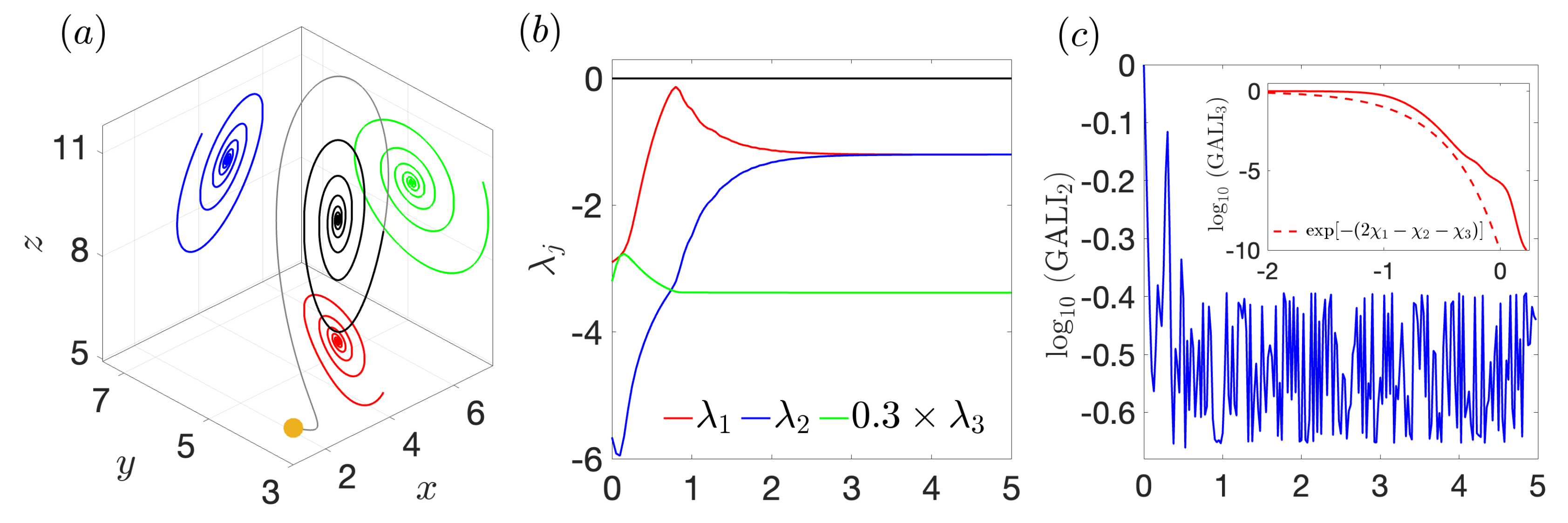}
	\includegraphics[width=1\columnwidth,keepaspectratio]{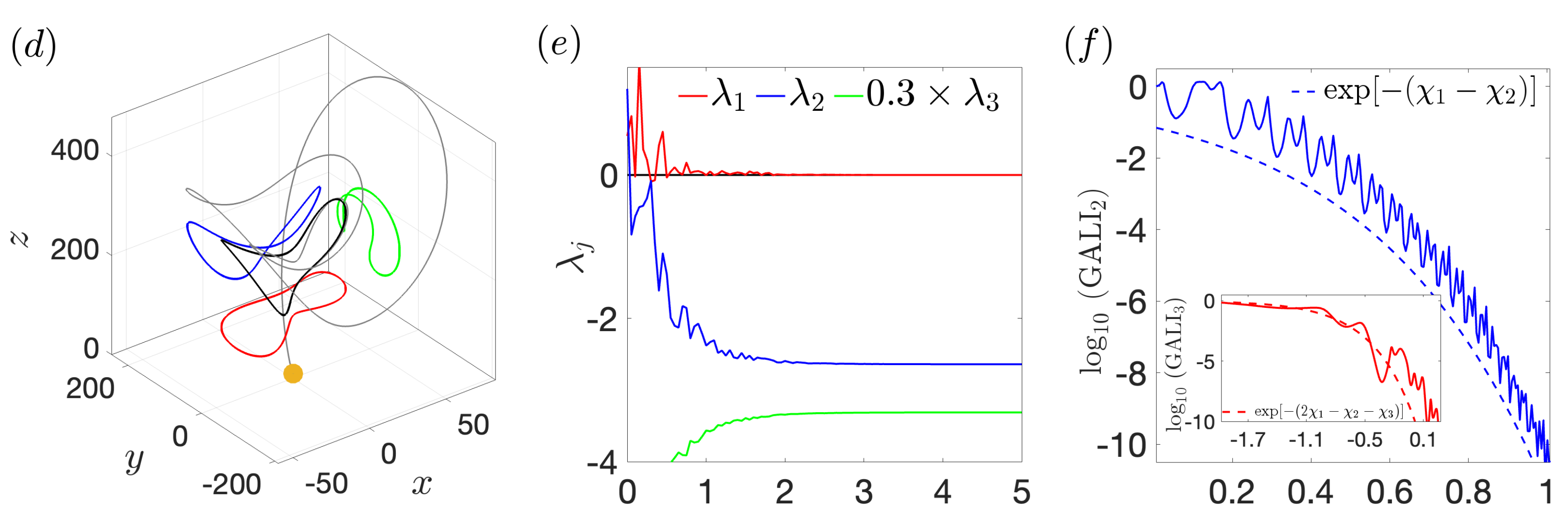}
  \vspace{-0.3cm}
	\includegraphics[width=1\columnwidth,keepaspectratio]{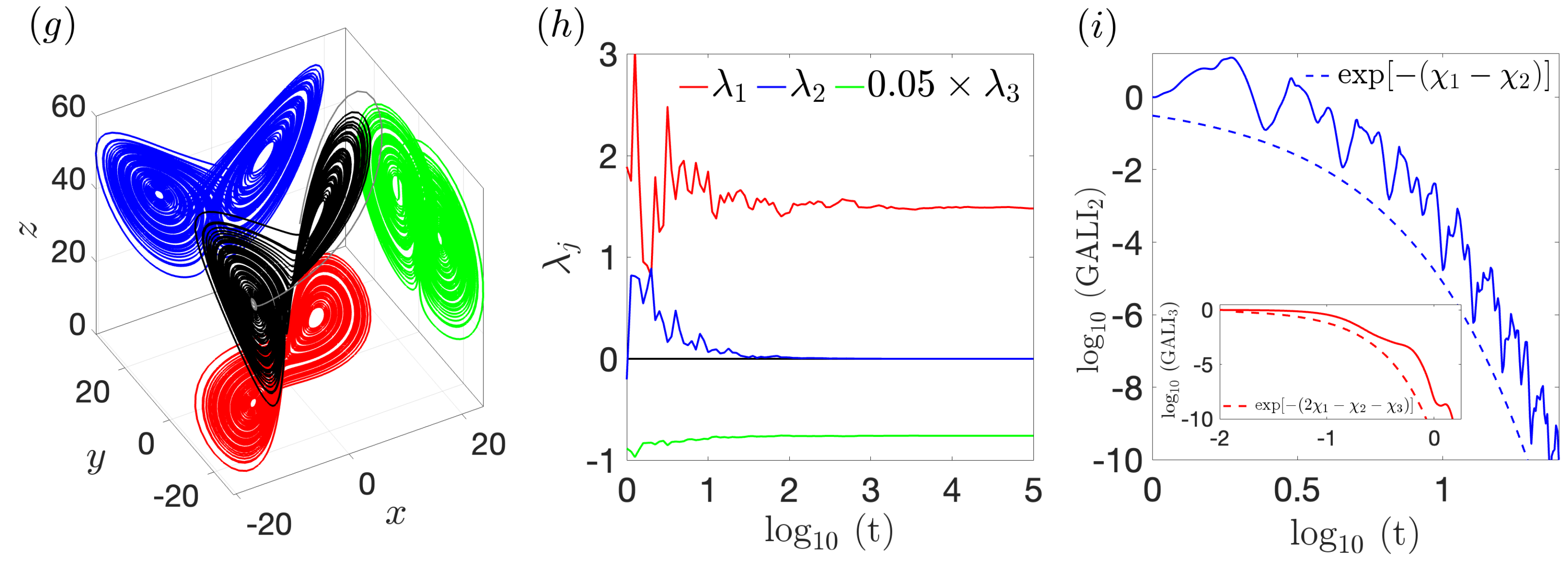}
 \vspace{-0.3cm}
\caption{(Left column) 3D phase space portraits of trajectories with IC $(x, y, z) = (1, 3, 6)$ [indicated by an orange circle point in (a) and (d), while in (g) is hidden behind the orbit]   for the 3D Lorenz system \eqref{eq:3DODE} with parameters $a=10$, $b=8/3$ and (a) $r=2.1$, (d)  $r=1$, and (g) $r=33.3$. The trajectory asymptotically tends to (a) a stable fixed point, (d) a stable limit cycle, and (g) a chaotic strange attractor. In gray  we depict the initial part of the trajectory's evolution and in black its asymptotic behavior, while in red, blue and green  we show its  2D  $xy$, $xz$, and $yz$ projections  respectively. (Middle column) The time evolution of the ftLEs of the trajectories depicted in the respective panel of the left column: $\lambda_1$ (red curves), $\lambda_2$ (blue curves), and $\lambda_3$ (green curves). The black line in each panel indicates $\lambda_j=0$ for comparison. Note that in all panels  the $\lambda_3$ values have been scaled for visualization purposes. (Right column) The time evolution of the GALI$_2$ (solid blue  curves) and the GALI$_3$ (solid red  curves in the inset plots) for the orbits depicted in the first panel of each row.  Apart from the GALI$_2$ in (c),  which oscillates around a constant positive value, all GALIs decay exponentially fast to zero, following the functional forms (dashed curves) given in \eqref{eq:GALI_chaos}  based on the LEs estimations obtained from the result presented in the plots of the middle column of panels. In particular, these values are: (c) $\chi_1=-1.20$, $\chi_2=-1.20$,  $\chi_3=-11.26$, (f) $\chi_1 = 0$, $\chi_2=-2.67$, $\chi_3=-11$, and (i) $\chi_1 =1.02$, $\chi_2 = 0$, and $\chi_3=-14.69$. 
}
	\label{fig:3D-01N}
\end{figure}

Figure \ref{fig:3D-01N}(a) illustrates the phase space $(x , y,z)$ portrait of the 3D Lorenz system \eqref{eq:3DODE} with parameters $a=10$, $b=8/3$, and $r=2.1$, for the orbit with IC $(x, y, z) = (1, 3, 6)$ (indicated by an orange circle point). The orbit's  evolution (represented by a gray and black curve) results in a trajectory tending to  a stable fixed point at $(x^*, y^*, z^*) = (4.899, 4.899, 9)$. We note that here, and throughout the paper, we use gray color to indicate the initial time interval of the trajectory's evolution, while we also show the different 2D projections of the trajectory on the planes $xy$ (red curve), $xz$ (blue curve), and $yz$ (green curve). In Fig.~\ref{fig:3D-01N}(b) we show the time evolution of the three ftLEs, $\lambda_j$, $j=1,2,3$, of the trajectory depicted in Fig.~\ref{fig:3D-01N}(a). As expected for a stable fixed point (see Sec.~\ref{sec:LEs}), all ftLEs remain negative. In particular, $\lambda_{1}$ (red curve) and $\lambda_{2}$ (blue curve) converge to almost identical values $\chi_1 \approx \chi_2 \approx -1.20$ at time $t \approx 3.5 \times 10^{4}$, while $\lambda_3$ (green curve) practically attains its asymptotic value $\chi_3=-11.26$ earlier at $t \approx 7.9 \times 10^{3}$. We note that in Fig.~\ref{fig:3D-01N}(b) we have scaled the $\lambda_3$ values for visualization purposes. In Fig.~\ref{fig:3D-01N}(c) we plot the time evolution of the GALI$_2$ (solid blue curve) and the GALI$_3$ (solid red curve, inset plot) of the same trajectory. This behavior of the GALI$_2$ is similar to what is observed for regular orbits in conservative Hamiltonian systems, and in accordance with the theoretical prediction of \eqref{eq:GALI_regular}. This similarity is not due to the dynamical resemblance between the two types of trajectories (actually, the asymptotic approach of a stable fixed point in the dissipative system \eqref{eq:3DODE} observed in Fig.~\ref{fig:3D-01N}(a), is quite different from the regular motion taking place on a torus in the phase space of a conservative Hamiltonian system), but is based on the fact that for both orbital behaviors the two largest LEs, $\chi_1$ and  $\chi_2$, are equal ($\chi_1 \approx \chi_2 \approx -1.20$ for the trajectory of Fig.~\ref{fig:3D-01N}(a), while $\chi_1 =\chi_2 =0$ for a regular orbit of a Hamiltonian system). Another way to understand the behavior of the GALI$_2$ in the case of orbits characterized by $\chi_1 \approx \chi_2$, is to notice that the more general behavior of the index given in \eqref{eq:GALI_chaos} results in the prediction of \eqref{eq:GALI_regular} for $k=2$ and $\chi_1 = \chi_2$. On the other hand, the GALI$_3$, shown in the  inset plot of Fig.~\ref{fig:3D-01N}(c), decays to zero following the exponential law $\exp {\left[-(2\chi_1 - \chi_2 - \chi_3) \right]}$, with $\chi_1=-1.20$, $\chi_2=-1.20$, and $\chi_3=-11.26$ (dashed red curve) in accordance with \eqref{eq:GALI_chaos}. 

\subsubsection{A stable limit cycle case}
\label{sec:3DODE_lc}

Figure \ref{fig:3D-01N}(d) depicts an example of a  stable limit cycle attractor occurring in the 3D Lorenz system \eqref{eq:3DODE} with $a=10$, $b=8/3$, and  $r=1$. In particular, we see the evolution of a trajectory with IC  $(x, y, z) = (1, 3, 6)$, which, as in Fig.~\ref{fig:3D-01N}(a), is represented by an orange circle point.  After an initial transit phase (gray curve), the orbit approaches the black colored stable limit circle, i.e., a closed phase space trajectory exhibiting periodic behavior. In Fig.~\ref{fig:3D-01N}(e) we show the  time evolution of the trajectory's three ftLEs. The largest one, $\lambda_1$, decreases asymptotically to zero, while $\lambda_2$ and $\lambda_3$ eventually remain constant having negative values, which can be considered as very good approximations of the orbit's LEs: $\chi_2=-2.67$ and $\chi_3=-11$. We note that in Fig.~\ref{fig:3D-01N}(e)  the values of $\lambda_3$ have been scaled for visualization purposes.  In Fig.~\ref{fig:3D-01N}(f) we present the time evolution of the trajectory's GALI$_2$ (solid blue curve) and GALI$_3$ (solid red curve, inset plot). Since the values of the LEs are not equal, both indices decay exponentially fast to zero with rates defined by the theoretical prediction \eqref{eq:GALI_chaos}, namely $\mbox{GALI}_2 \propto \exp {\left[ -(\chi_1 - \chi_2) \right]}$ and $\mbox{GALI}_3 \propto \exp {\left[ -(2\chi_1 - \chi_2 - \chi_3) \right]}$ (dashed curves), with $\chi_1 = 0$, $\chi_2=-2.67$, $\chi_3=-11$.

\subsubsection{A chaotic, strange attractor case}
\label{sec:3DODE_sa}
 
In order to obtain a typical strange attractor in the 3D Lorenz system \eqref{eq:3DODE}, we keep the same IC as before, i.e., $(x, y, z) = (1, 3, 6)$, and set the system's parameters to  $a=10$, $b=8/3$, and  $r=33.3$. In Fig.~\ref{fig:3D-01N}(g) we see  the respective phase portrait of this trajectory (although the IC is not visible as is hidden behind the orbit), while,  as is shown in  Fig.~\ref{fig:3D-01N}(h), for this orbit we eventually get $\lambda_1 > 0$, $\lambda_2 = 0$ and $\lambda_{3} < 0$. The evolution of the three ftLEs in  Fig.~\ref{fig:3D-01N}(h) allows us to estimate  the LEs as $\chi_1 =1.02$, $\chi_2 = 0$, and $\chi_3=-14.69$. Since, as in the case of the limit cycle studied in Sec.~\ref{sec:3DODE_lc}, the LEs have distinct values, the GALI$_{2}$ [solid blue curve in Fig.~\ref{fig:3D-01N}(i)] and GALI$_{3}$ [solid red curve in the inset of Fig.~\ref{fig:3D-01N}(i)] indices decay exponentially fast to zero, following functions proportional to $\exp {\left[ -(\chi_1 - \chi_2) \right]}$ and $\exp {\left[ -(2\chi_1 - \chi_2 - \chi_3) \right]}$ respectively [dashed curves in Fig.~\ref{fig:3D-01N}(i)], in accordance with \eqref{eq:GALI_chaos}.

\subsubsection{Parametric exploration of the 3D Lorenz system's  dynamics}
\label{sec:3DODE_param}
 
So far, we have examined a small number of exemplary trajectories of the 3D Lorenz system \eqref{eq:3DODE}, by choosing different values for the model parameter $r$. We now perform a more global investigation of the GALI$_2$ performance, in comparison to the behavior of the system's LEs, for a range of $r$ values, while the other parameters are kept fixed to $a=10$ and $b=8/3$. Since the dynamical classification of the different types of trajectories  in Sec.~\ref{sec:LEs} was done with respect to the orbit's  LEs spectrum, we first conduct a parameter exploration of this spectrum and then we compare our findings with the obtained GALI$_2$ results. In our analysis we consider only the GALI$_2$ method, because the  GALI$_3$, as was also shown in Figs.~\ref{fig:3D-01N}(c), \ref{fig:3D-01N}(f) and \ref{fig:3D-01N}(i), decays exponentially fast to zero for all considered cases, and hence it does not provide any insight into the  distinction between different types of motion. We  note that for our investigations the ftLEs and the GALI$_2$ were computed for $t = 10^4$ time units, and that whenever GALI$_2 \leq 10^{-8}$ we stopped our calculations in order to reduce the required computational cost, considering this threshold value as a good indication that the index was practically equal to zero. In this section we present results  obtained  for the orbit with IC $(x, y, z) = (1, 3, 6)$, which is a good choice to study  the system's dynamical behaviors, as other ICs  practically gave similar results. 

\begin{figure}[h!]\centering
	\includegraphics[width=1\columnwidth,keepaspectratio]{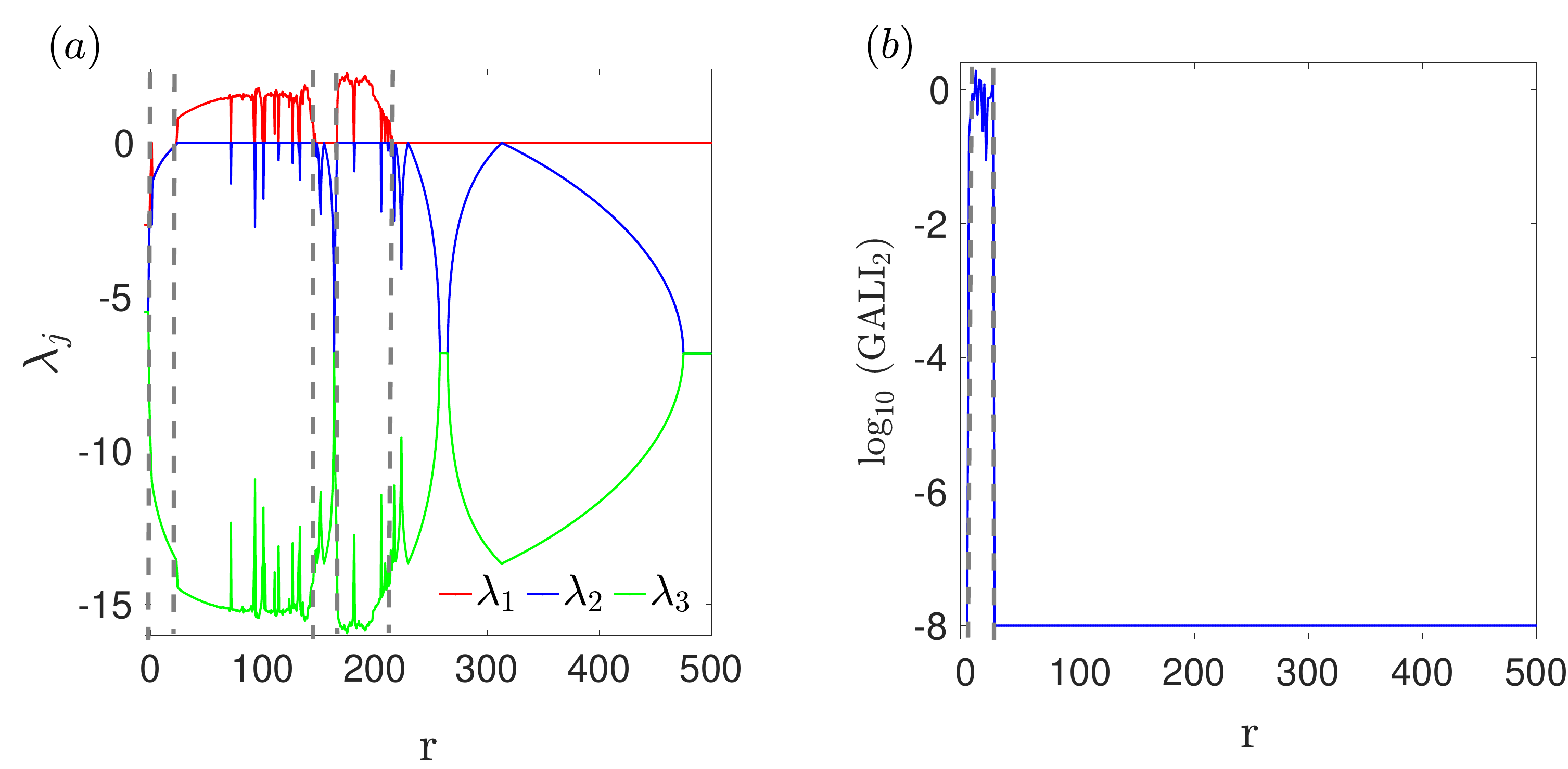} 
	\caption{The values, at $t=10^4$, of (a) the spectrum of the ftLEs $\lambda_1$,  $\lambda_2$, $\lambda_3$ (respectively depicted by red, blue, and green curves),  and (b) the GALI$_2$,   as a function of $r$ ($r \in [-5,500]$)  for the trajectory with IC $(x, y, z) = (1, 3, 6)$ of the  3D Lorenz system   \eqref{eq:3DODE} with $a=10$ and $b=8/3$. Gray vertical dashed lines indicate the values $r=1.3$, $21.3$, $146.9$, $166$, $215.4$ in (a),  and  the values $r=1.3$ and $21.3$ in (b).  }
\label{fig:3D-02}
\end{figure}

In Fig.~\ref{fig:3D-02}(a) we present the values of the three ftLEs, $\lambda_1$ (red curve),  $\lambda_2$ (blue curve), $\lambda_3$ (green curve) at $t=10^4$ (which can be considered as good approximations of the actual LEs) for $1,001$ equally distributed $r$ values in the range $-5 \leq r \leq 500$. For $- 5 \leq r \leq 21.3$ the system exhibits stable fixed points characterized by all ftLEs eventually being   negative.  The GALI$_2$ decays exponentially fast to zero, quickly reaching very small values  for  $-5 \leq r \leq 1.3$ [Fig.~\ref{fig:3D-02}(b)] where stable fixed points with $\lambda_1$, $\lambda_2$ having negative, but different values, occur. This behavior is due to the fact that the GALI$_2$ follows the time evolution described in \eqref{eq:GALI_chaos}, for which $\lambda_1 > \lambda_2$ leads to exponential decay. For  values in the interval $1.3 \leq r \leq  21.3$,  where $\lambda_1 \approx  \lambda_2$,  the GALI$_2$   asymptotically attains positive values,  in agreement with the prediction provided from  \eqref{eq:GALI_chaos}. From the results of Fig.~\ref{fig:3D-02}(a) we see that for $21.3<r\leq 146.9$ and $166 <r\leq 215.4$ the dynamics of the system is characterized by the presence of strange attractors ($\lambda_1>0$),   while the appearance of stable limit cycles  (characterized by $\lambda_1 =0$) is observed for  $146.9 < r \leq 166$ and $r > 215.4$.  For all of these cases, since $\lambda_1 \neq \lambda_2$,  the GALI$_2$ decays exponentially fast to zero as \eqref{eq:GALI_chaos} denotes. Thus, we understand that the index cannot differentiate between the various dynamical behaviors appearing when $\lambda_1 > \lambda_2$.
 
\begin{figure}[h!]\centering
\includegraphics[width=1.\columnwidth,keepaspectratio]{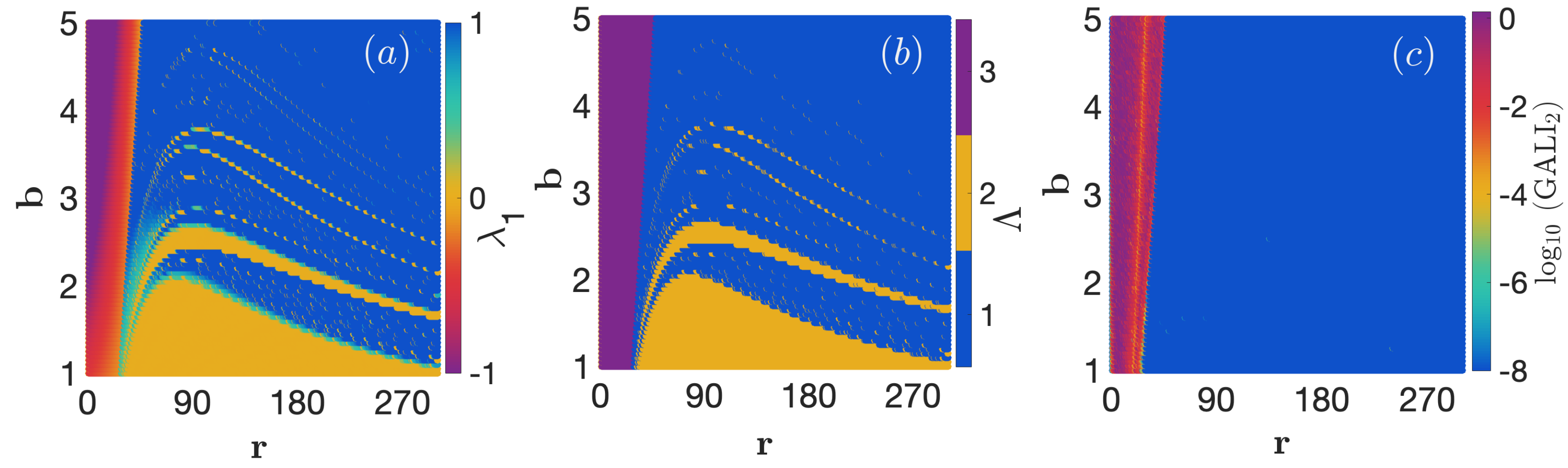}
\caption{The parameter space $(r,b)$  of the 3D Lorenz system \eqref{eq:3DODE} with $a=35$,  colored according to the value of (a) the ftmLE $\lambda_1$ (scaled in the interval $[-1,1]$), (b) the index $\Lambda$, and (c) the GALI$_2$ of the trajectory with  IC $(x, y, z) = (2, 1, 5)$, at $t=10^4$. In (b)  the index  $\Lambda$ is  $\Lambda=1$ when $\lambda_1>0$, $\lambda_2\leq 0$, $\lambda_3<0$ (blue region), indicating the presence of chaotic attractors, $\Lambda=2$ for $\lambda_1 \approx 0$, $\lambda_2 < 0$, $\lambda_3<0$ (orange region) denoting the existence of limit cycles, and $\Lambda=3$ when $\lambda_1, \lambda_2, \lambda_3<0$ (purple region) corresponding to the appearance of stable fixed points.  Each color plot is created by considering  a set of $2,991 \times 81 = 242,271$ equally spaced grid points on the region  $(r, b) = [0, 300] \times [1, 5]$.
 }
\label{fig:3D-Bipar}
\end{figure}

In Fig.~\ref{fig:3D-Bipar} we perform a  parametric exploration of the dynamics of the  3D Lorenz system \eqref{eq:3DODE} by varying both the $r$ and $b$ parameters, while keeping  the other parameter of the system fixed to $a=35$, as well as the IC of the considered trajectory at $(x, y, z) = (2, 1, 5)$.  For each combination of $r$ and $b$ we integrate the trajectory for $t=10^4$  time units and register the values of the ftLEs,  as well as of the GALI$_2$ at the end of the integration. We note that, in order to obtain the results of Fig.~\ref{fig:3D-Bipar} we changed the value of $a$,  along with the coordinates of the considered IC, with respect to what we used before,  solely for exploring different setups of the system. 

In Fig.~\ref{fig:3D-Bipar}(a) we color each point of the parameter space $(r,b)$ according to the final $\lambda_1$ value of the considered trajectory. To enhance visualization we implement a scaling approach on the computed $\lambda_1$ values, transforming them into the range $[-1,1]$ to emphasize their sign and closeness to zero. In particular, we map positive $\lambda_1$ values (in this  case we had $0 \leq \lambda_1 \leq 5.99$) to $[0,1]$, and negative values (the actual values were $-2.45 \leq \lambda_1 \leq 0$) to $[-1,0]$. This approach enables us to distinguish whether $\lambda_1$ is negative, zero or positive without focusing on its actual value. 

Thus, by identifying whether $\lambda_1<0$,  $\lambda_1 \approx 0$ or $\lambda_1>0$ we characterize the considered trajectory as tending to  a stable fixed point, a stable limit cycle or a chaotic attractor, respectively. Consequently, blue regions ($\lambda_1 > 0$) in Fig.~\ref{fig:3D-Bipar}(a) denote the existence of chaotic attractors, yellow/orange areas ($\lambda_1 \approx 0$)  represent parameter regions where stable limit cycles exist, while dark red colors ($\lambda_1 < 0$) denote the appearance of stable fixed points. It is worth noting that for the considered parameter ranges we have not detected motion on a stable $k$D torus, $k>1$, which would be characterized by having the $k$ largest LEs practically equal to zero.

To better categorize the different types of  dynamical behaviors appearing in the 3D Lorenz system \eqref{eq:3DODE} we perform a classification based on the values of the whole spectrum of LEs. To do that we use an index $\Lambda$ to denote the various observed arrangements. In particular, we set $\Lambda=1$ when the combination $\lambda_1>0$, $\lambda_2\leq 0$, $\lambda_3<0$, corresponding to the presence of chaotic attractors, is observed, while $\Lambda=2$ denotes the presence of a limit cycle  (i.e., $\lambda_1 \approx 0$, $\lambda_2 < 0$, $\lambda_3<0$), and $\Lambda=3$ corresponds to $\lambda_i <0$, $i=1,2,3$, indicating the existence of stable fixed points. The outcome of this process is depicted in Fig.~\ref{fig:3D-Bipar}(b), where the parameter values associated with different asymptotic dynamical behaviors are denoted by diverse colors: blue ($\Lambda=1$, chaotic attractors), orange ($\Lambda=2$, stable limit cycles), and purple ($\Lambda=3$, stable fixed points). By comparing Figs.~\ref{fig:3D-Bipar}(a) and \ref{fig:3D-Bipar}(b)  we see that both of them capture, in a practically similar way, the dynamical behavior of the system, by clearly identifying parameter regions where the studied trajectories eventually tend to different attractors. The separation between the different dynamical regions is clearer in Fig.~\ref{fig:3D-Bipar}(b) because the coloring of the parameter space is not continuous, as only three colors are used. Nevertheless, the identifications of the different types of motions is also very well done by computing only the  ftmLE in  Fig.~\ref{fig:3D-Bipar}(a), something which is computationally easier than finding the values of the whole spectrum of LEs needed for creating  Fig.~\ref{fig:3D-Bipar}(b). Thus, we conclude that using only the value of $\lambda_1$ is sufficient to reveal the different dynamical behaviors exhibited by the 3D Lorenz system \eqref{eq:3DODE} for the parameter ranges considered in Fig.~\ref{fig:3D-Bipar}.

In Fig.~\ref{fig:3D-Bipar}(c) we present a similar analysis to the one performed in Fig.~\ref{fig:3D-Bipar}(a), but by using the GALI$_2$ values. From the results of this figure we see that the GALI$_2$ becomes practically zero for the majority of the considered cases, as the largest part of the parametric  space is colored in blue. This is due to the exponential decay of the GALI$_2$ to zero, in accordance with \eqref{eq:GALI_chaos},  as in all these cases $\lambda_1 \neq \lambda_2$. Only in the leftmost region of Fig.~\ref{fig:3D-Bipar}(c) (small $r$ values),  where stable fixed point attractors exist according to Figs.~\ref{fig:3D-Bipar}(a) and \ref{fig:3D-Bipar}(b), the GALI$_2$ attains nonzero positive values (region colored in purple/red) in accordance with the predictions of \eqref{eq:GALI_chaos}, as in these cases the two largest LEs have negative,  but almost identical values.

It is clear from  Fig.~\ref{fig:3D-Bipar} that the computation of the spectrum of LEs [Fig.~\ref{fig:3D-Bipar}(b)], or even  the estimation of only the mLE [Fig.~\ref{fig:3D-Bipar}(a)], manages to capture the existence of different types of attractors in the phase space of the   3D Lorenz system \eqref{eq:3DODE}. On the other hand,  the computation of the GALI$_2$ succeeds in identifying the presence of only stable fixed points, which are characterized by $\lambda_1 \approx \lambda_2$,  but fails to discriminate between other dynamical behaviors for which $\lambda_1 > \lambda_2$, as in all these cases it becomes zero following the exponential decay of \eqref{eq:GALI_chaos}.

\subsection{Numerical investigation of the 4D Lorenz hyperchaotic system} 
\label{sec:4DODE}

To further explore the behavior of the GALI method for hyperchaotic attractors, we employ the 4D Lorenz model \eqref{eq:4DODE}. The addition of a fourth dimension allows the system to attain two positive LEs and more complex dynamics compared to its 3D counterpart model \eqref{eq:3DODE}. In this section, we perform a similar analysis to the one presented  in Sec.~\ref{sec:3DODE}.

\subsubsection{A stable fixed point case}
\label{sec:4DODE_fp}

\begin{figure}[ht]\centering
\includegraphics[width=1.\columnwidth,keepaspectratio]{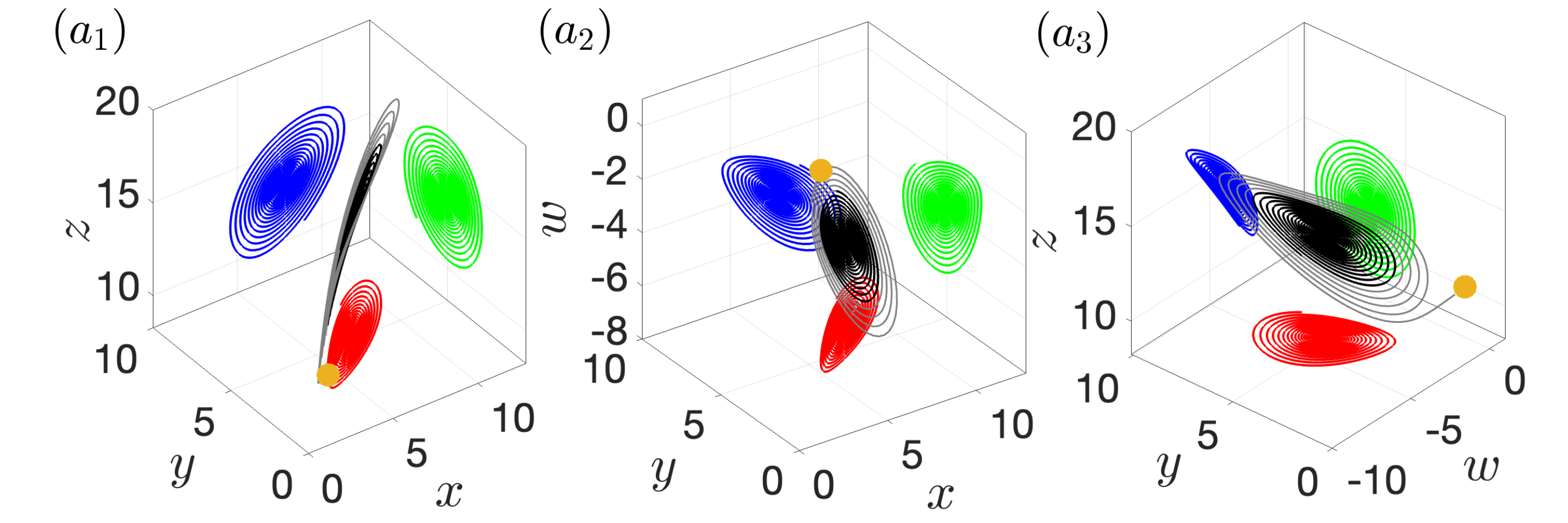}
\includegraphics[width=1.\columnwidth,keepaspectratio]{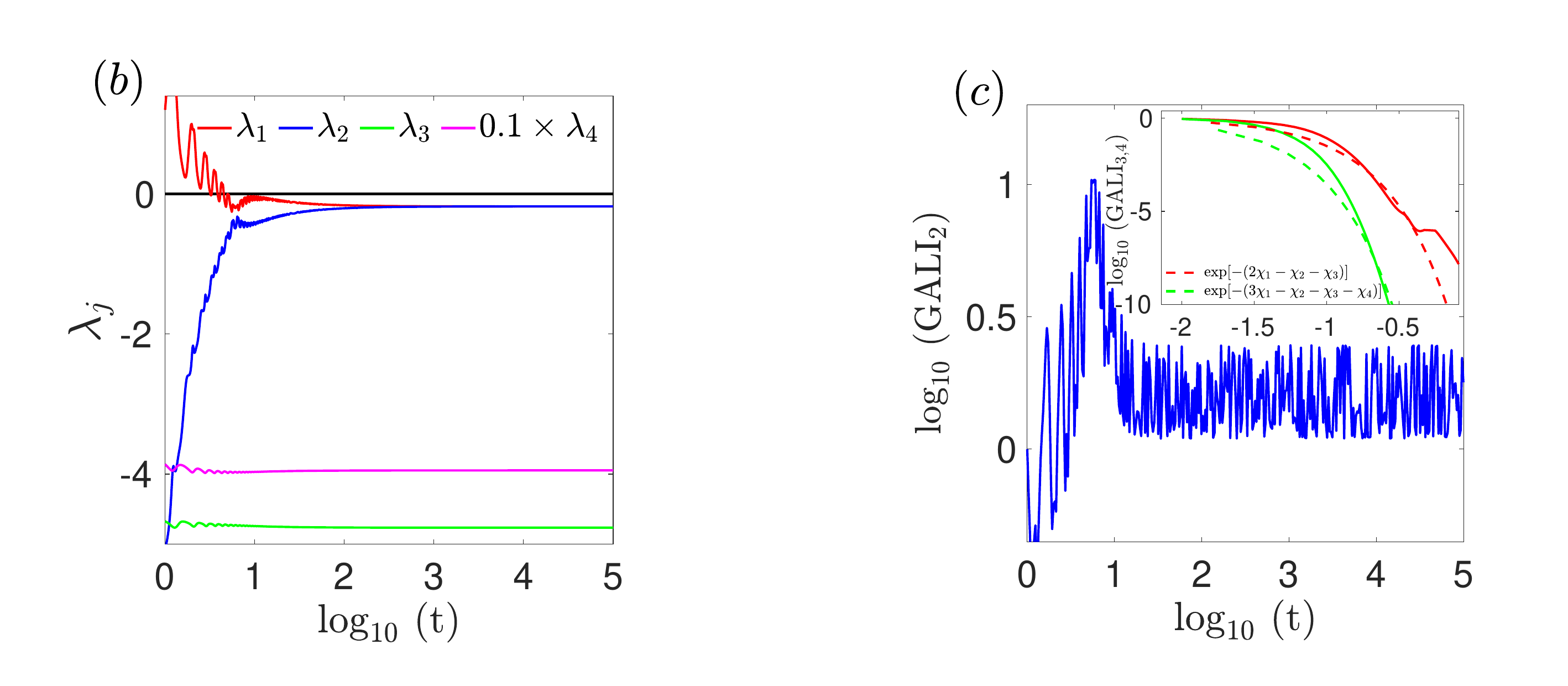}
\caption{[(a$_1$), (a$_2$), (a$_3$)] $3$D phase space projections of the trajectory with IC $(x,y,z,w) = (3,2,10,1)$ (orange circle points), which asymptotically approaches a stable fixed point attractor of the $4$D Lorenz system \eqref{eq:4DODE} with $a=35$, $b=8/3$, $c=2$, and $r=-12$. As  in the  left column panels of Fig.~\ref{fig:3D-01N}, we use gray color to depict the initial phase of the trajectory’s evolution, and red, green and blue colors to show  projections of the orbit in  different  2D planes. (b) The time evolution of the four ftLEs of the trajectory. The $\lambda_4$ values  have been rescaled for visualization purposes. The horizontal black line  indicates  $\lambda_j=0$  for  comparison. (c) The time evolution of the GALI$_{2}$ (solid blue curve) displays  fluctuations around a constant positive value due to the fact that $\lambda_1$ and $\lambda_2$ become practically equal. The GALI$_3$ (solid red curve) and the GALI$_4$ (solid green curve) in the inset, decay  to zero following specific exponential laws provided by \eqref{eq:GALI_chaos} (dashed curves). }
\label{fig:4D-01}
\end{figure}

We begin our investigation by considering the trajectory with IC $(x,y,z,w) = (3,2,10,1)$ of the 4D Lorenz system \eqref{eq:4DODE} with parameters $a=35$, $b=8/3$, $c=2$, and $r=-12$, which results in an orbit asymptotically approaching  a stable fixed point attractor. Since  system \eqref{eq:4DODE} involves four state variables, $x, y, z$ and $w$, the direct visualization of the entire phase space  is challenging, so we plot all possible $3$D projections of the $4$D space, i.e., $(x,y,z)$, $(x,y,w)$, and $(y,z,w)$, in Figs.~\ref{fig:4D-01}(a$_1$)$-$(a$_3$) respectively. We note that in each plot the trajectory's IC is denoted by an orange circle point. For each 3D phase space plot we  also depict the 2D projections of the trajectory in the respective planes, similarly to what was done in the left column panels of Fig.~\ref{fig:3D-01N}. In Figs.~\ref{fig:4D-01}(a$_1$)$-$(a$_3$) we clearly see  that this trajectory  eventually tends to  the stable fixed point attractor $(x^*, y^*, z^*, w^*) = (7.141, 5.129, 13.733, -3.923)$.  All the LEs of the trajectory have negative values [Fig.~\ref{fig:4D-01}(b)], as expected for stable fixed point attractors, reflecting the system's phase space contraction. In addition, the  two largest ftLEs, $\lambda_1$ and $\lambda_2$, eventually saturate to the same negative value, indicating that the system exhibits uniform contraction along the related directions. Due to the practical equality of $\lambda_1$ and $\lambda_2$, and in agreement with the theoretical prediction \eqref{eq:GALI_chaos}, the trajectory's GALI$_{2}$ [solid blue  curve in Fig.~\ref{fig:4D-01}(c)] fluctuates around a constant positive value. On the other hand, the GALI$_{3}$ and the  GALI$_4$ indices [red and green solid curves in the inset of Fig.~\ref{fig:4D-01}(c) respectively] exhibit  exponential decays. The good agreement between the actual GALI$_3$ and GALI$_4$ values, and the theoretical expectation provided by \eqref{eq:GALI_chaos}, is verified by the proximity of the red and green dashed curves in the inset of Fig.~\ref{fig:4D-01}, which  denote functions proportional to $\exp {\left[ -(2\chi_1 - \chi_2 - \chi_3) \right]}$ and $\exp {\left[ -(3\chi_1 - \chi_2 - \chi_3 - \chi_4) \right]}$ respectively. These functions are computed for $\chi_1 = -1.23$, $\chi_2=-1.23$, $\chi_3=-11.94$, and $\chi_4 = -50.66$, which are good estimations of the trajectory's LEs, obtained from the results of Fig.~\ref{fig:4D-01}(b).
\FloatBarrier

\subsubsection{A stable limit cycle case}
\label{sec:4DODE_lc}

\begin{figure}[ht]\centering
	\includegraphics[width=1.\columnwidth,keepaspectratio]{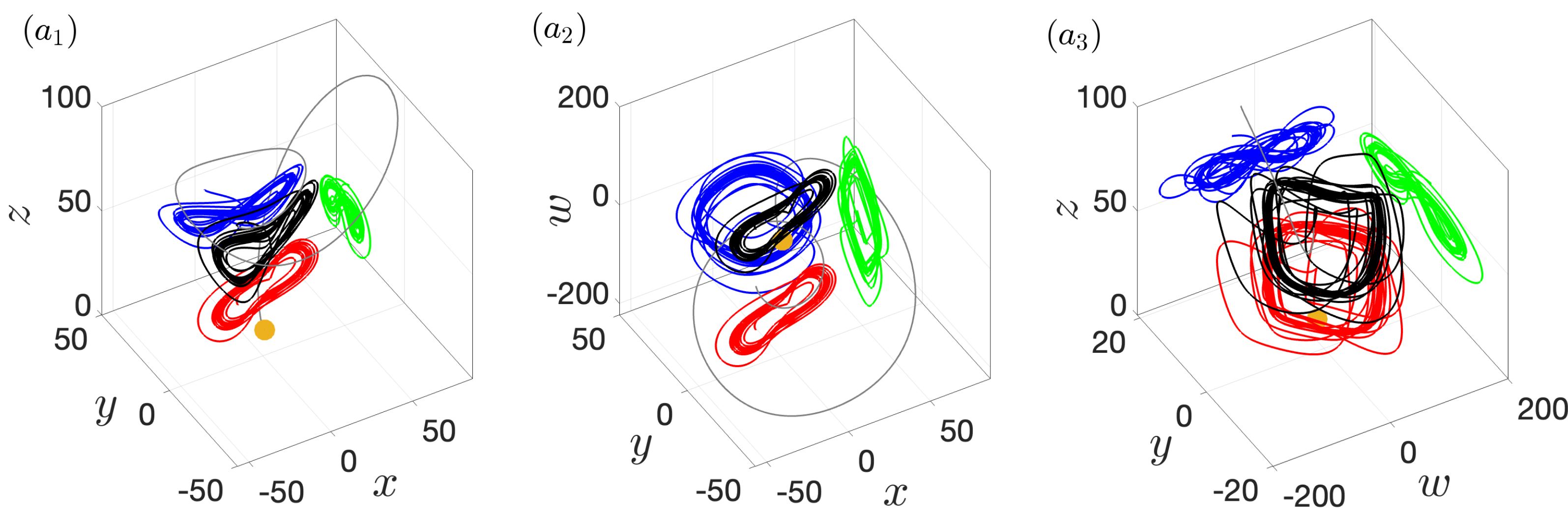}
	\includegraphics[width=1.\columnwidth,keepaspectratio]{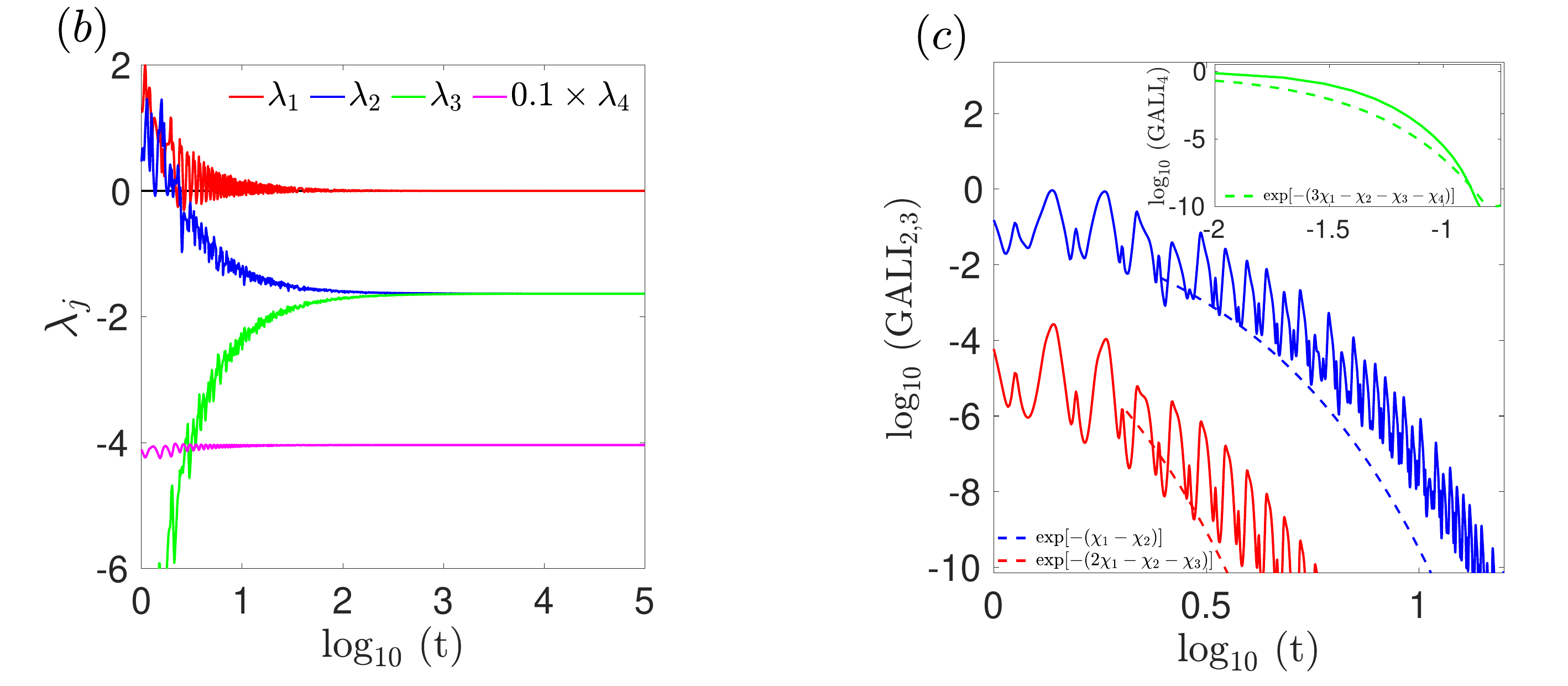}
\caption{[(a$_1$), (a$_2$), (a$_3$)] $3$D phase space projections of the trajectory with IC $(x,y,z,w) = (3,2,10,1)$ (orange circle points),  which asymptotically approaches a stable limit cycle of the $4$D Lorenz system \eqref{eq:4DODE} with $a=35$, $b=8/3$, $c=2$, and $r=-5$. As in   Fig.~\ref{fig:4D-01}, we use gray color to depict the initial phase of the trajectory’s evolution, and red, green and blue colors to show  projections of the orbit in  different  2D planes. (b) The time evolution of the four ftLEs of the trajectory. The $\lambda_4$ values  have been rescaled for visualization purposes. The horizontal black line (not clearly seen due to the overlap of the $\lambda_1$ values) indicates  $\lambda_j=0$  for  comparison. (c) The  GALI$_{2}$ (solid blue curve),   GALI$_3$ (solid red curve) and  GALI$_4$ (solid green curve in the inset) decay  to zero following specific exponential laws provided by \eqref{eq:GALI_chaos} (dashed curves).
}
        \label{fig:4D-02}
\end{figure}

To study a case of a stable limit cycle in the 4D Lorenz system \eqref{eq:4DODE}, we set $r=-5$, while keeping all the other parameters and the considered IC as  in Sec.~\ref{sec:4DODE_fp}, namely $a=35$, $b=8/3$, $c=2$ and $(x,y,z,w) = (3,2,10,1)$.
In  Figs.~\ref{fig:4D-02}(a$_1$)$-$(a$_3$), where we present the 3D projections of this trajectory,  we see that after an initial transient phase (colored in gray), the trajectory becomes confined on a closed loop in all 3D projections (and consequently in the related 2D projections). The trajectory's ftmLE, $\lambda_1$, converges to zero, while the remaining ftLEs tend to  negative values [Fig.~\ref{fig:4D-02}(b)], which according to the classification presented in Sec.~\ref{sec:LEs}, indicates that the system exhibits a stable limit cycle. Since $\lambda_1 > \lambda_2$, all  GALI$_{k}$, $k=2,3,4$, indices [solid blue, red and green curves respectively in  Fig.~\ref{fig:4D-02}(c) and its inset]
decay to zero exponentially fast following evolutions  proportional to $\exp {\left[ -(\chi_1 - \chi_2) \right]}$, $\exp {\left[ -(2\chi_1 - \chi_2 - \chi_3) \right]}$, and $\exp {\left[ -(3\chi_1 - \chi_2 - \chi_3 - \chi_4) \right]}$ respectively, with $\chi_1 = 0$, $\chi_2=-1.63$, $\chi_3=-1.63$, and $\chi_4 = -40.37$ [denoted by dashed curves in Fig.~\ref{fig:4D-02}(c)], in agreement with \eqref{eq:GALI_chaos}.
\FloatBarrier

\subsubsection{A chaotic, strange attractor case}
\label{sec:4DODE_sa}

\begin{figure}[htbp]\centering
	\includegraphics[width=1.\columnwidth,keepaspectratio]{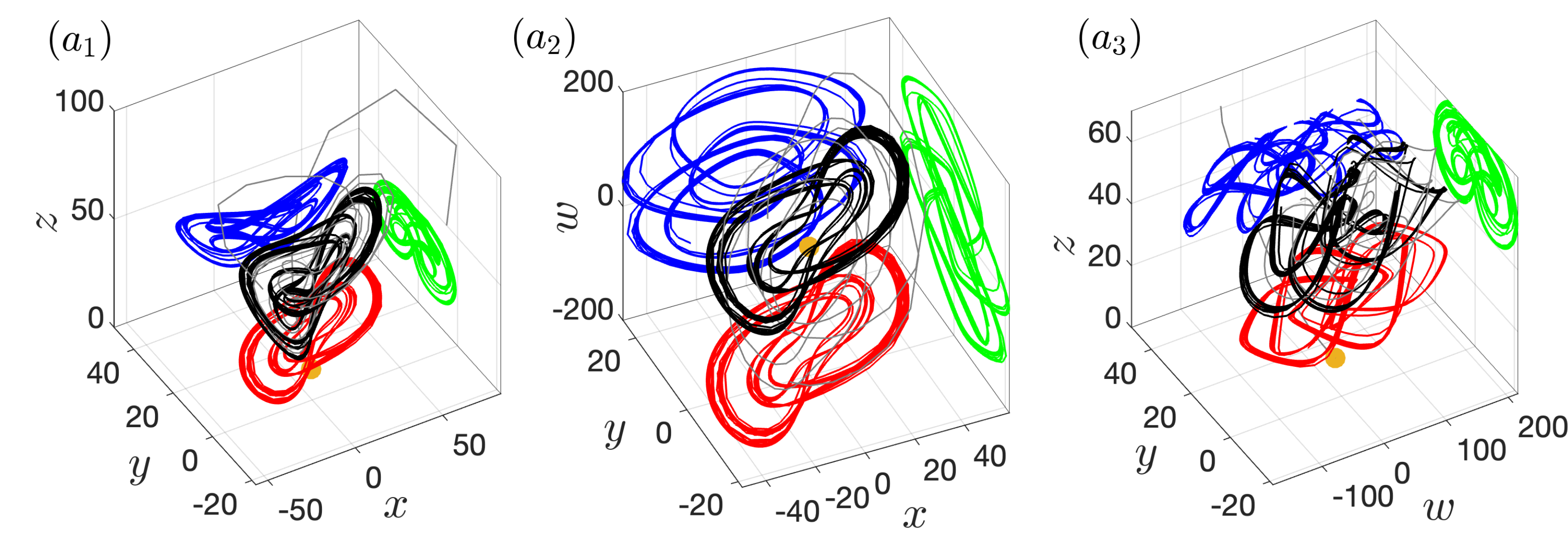}
	\includegraphics[width=1.\columnwidth,keepaspectratio]{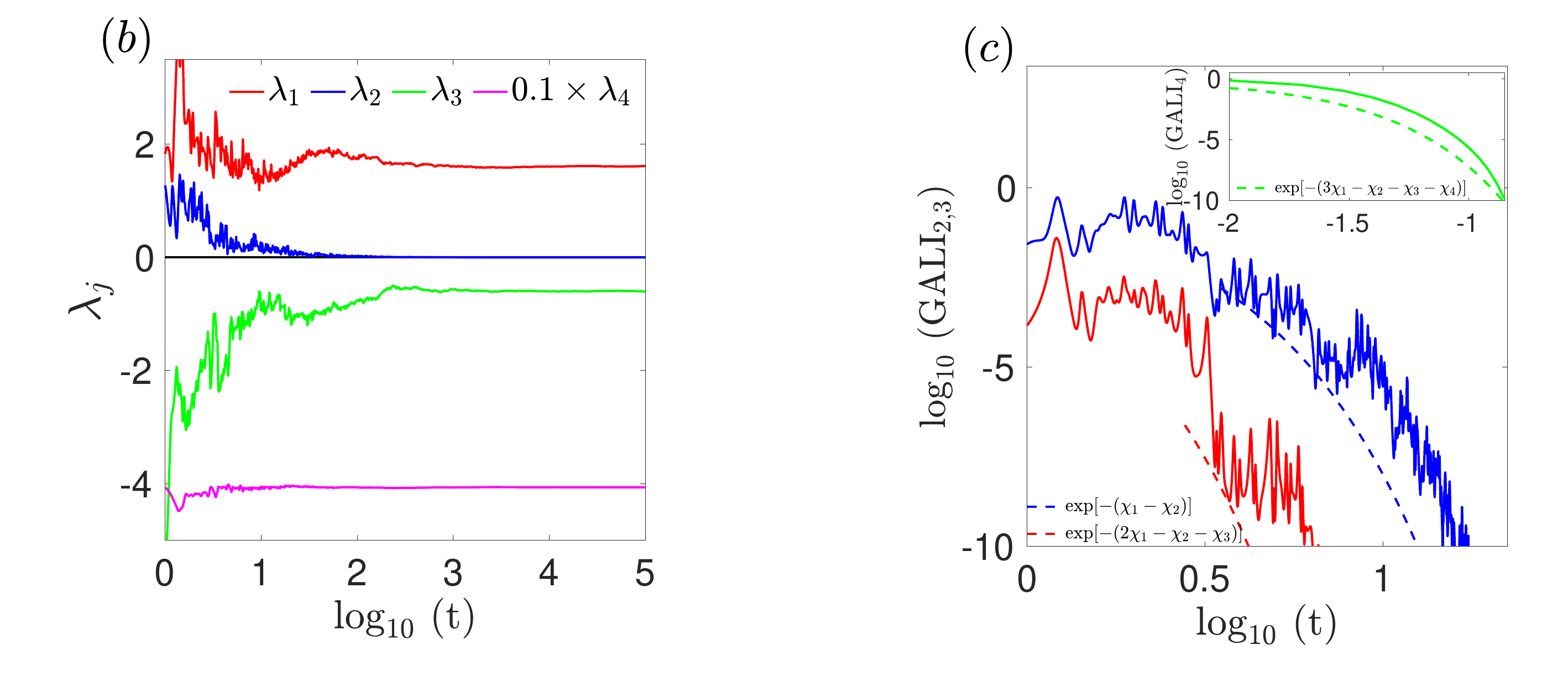}
\caption{Similar to Fig.~\ref{fig:4D-02}, but for the trajectory with IC $(x,y,z,w) = (3,2,10,1)$  of the 4D Lorenz system \eqref{eq:4DODE} with $a=35$, $b=8/3$, $c=55$ and $r=-1$. The trajectory tends to a chaotic attractor having $\chi_1 = 1.60$, $\chi_2=0$, $\chi_3=-0.59$, and $\chi_4 = -40.64$.  }
\label{fig:4D-03}
 \end{figure}

By setting $r=1.5$, while keeping $a=35$, $b=8/3$, $c=2$ and the IC $(x,y,z,w) = (3,2,10,1)$ as before, we obtain for system \eqref{eq:4DODE} a trajectory which tends to  a chaotic attractor. In  Figs.~\ref{fig:4D-03}(a$_1$)$-$(a$_3$) we show the 3D phase space projections  of this  trajectory, which exhibits a rather complex behavior on a  chaotic  attractor. From the trajectory's  LEs, only $\lambda_{1}$ remains positive, tending to a constant number $\chi_1 = 1.60$, while $\lambda_{2}$ becomes zero, and $\lambda_{3}$ and $\lambda_{4}$ eventually attain negative values indicating that $\chi_3=-0.59$, and $\chi_4 = -40.64$ [Fig.~\ref{fig:4D-03}(b)]. In Fig.~\ref{fig:4D-03}(c) and its inset, the exponential decay of the GALI$_2$ (solid blue curve), GALI$_3$ (solid red curve) and   GALI$_4$ (solid green curve) is clearly seen. These decays are well approximated by GALI$_2 \propto \exp {\left[ -(\chi_1 - \chi_2) \right]}$, GALI$_3 \propto \exp {\left[ -(2\chi_1 - \chi_2 - \chi_3) \right]}$, and GALI$_4 \propto \exp {\left[ -(3\chi_1 - \chi_2 - \chi_3 - \chi_4) \right]}$ (dashed curves) for $\chi_1 = 1.60$, $\chi_2=0$, $\chi_3=-0.59$ and $\chi_4 = -40.64$.
\FloatBarrier

\subsubsection{A hyperchaotic attractor case}
\label{sec:4DODE_hyper}
 
\begin{figure}[htbp]\centering
	\includegraphics[width=1.\columnwidth,keepaspectratio]{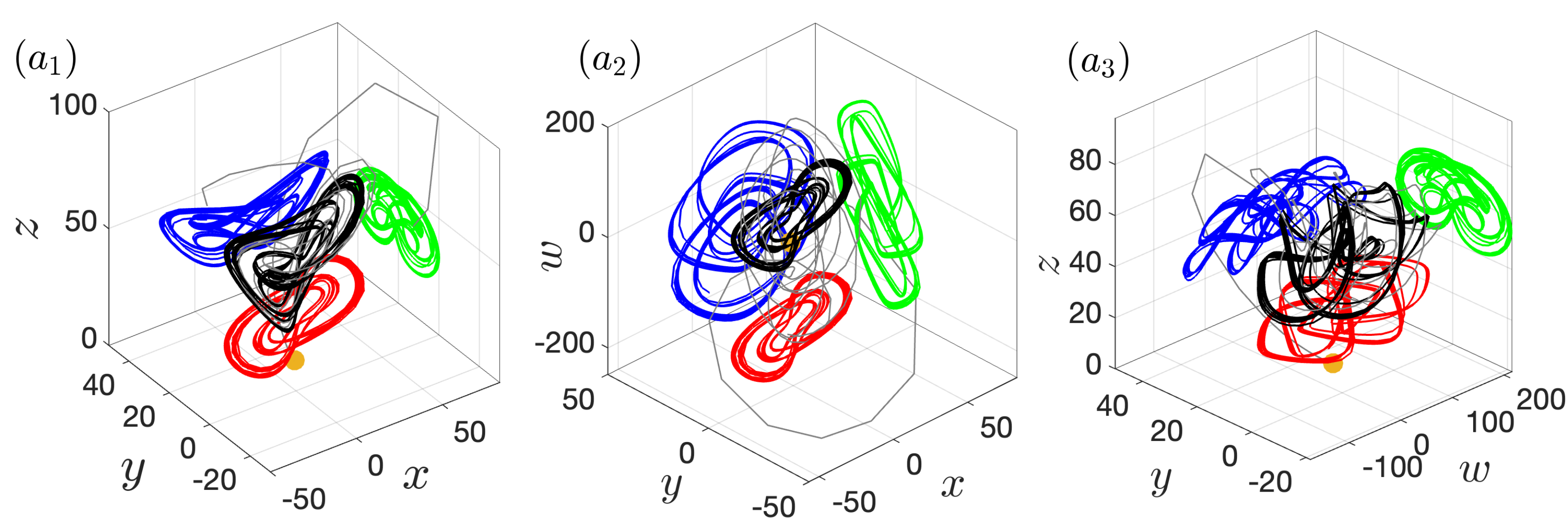}
	\includegraphics[width=1.\columnwidth,keepaspectratio]{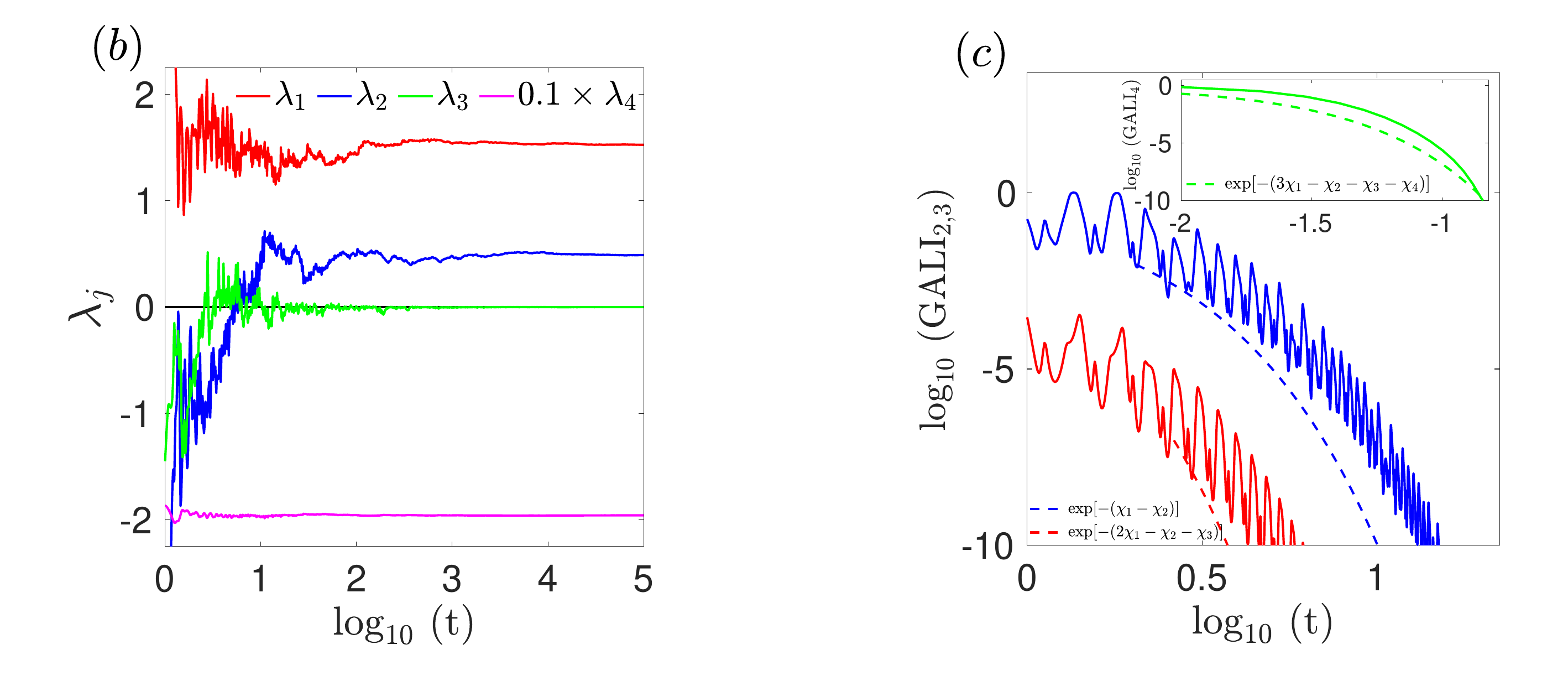}
	\caption{Similar to Fig.~\ref{fig:4D-02}, but for the trajectory with IC $(x,y,z,w) = (3,2,10,1)$  of the 4D Lorenz system \eqref{eq:4DODE} with $a=35$, $b=8/3$, $c=55$  and $r=1.5$. The trajectory exhibits hyperchaotic behavior  having  $\chi_1 = 1.53$,  $\chi_2=0.51$ $\chi_3=0$  and $\chi_4 = -39.19$. } 
	\label{fig:4D-04}
\end{figure}

The last case we consider in  the 4D Lorenz system \eqref{eq:4DODE} is a trajectory with IC $(x,y,z,w) = (3,2,10,1)$  for  $a=35$, $b=8/3$, $c=55$   and $r=1.5$, which exhibits hyperchaotic behavior. The various 3D phase space projections of this trajectory are shown in Figs.~\ref{fig:4D-04}(a$_1$)$-$(a$_3$),  and the time evolution of its ftLEs in Fig.~\ref{fig:4D-04}(b). The hyperchaotic nature of the dynamics is reflected on the attained positive values of the two largest ftLEs, whose evolution in Fig.~\ref{fig:4D-04}(b) suggests that the corresponding LEs are $\chi_1 = 1.53$ and $\chi_2=0.51$. The other two ftLEs also tend to constant values indicating that $\chi_3=0$  and $\chi_4 = -39.19$. As  expected from the theoretical prediction \eqref{eq:GALI_chaos}, and since the two largest LEs are different from each other,  all GALIs tend exponentially fast to zero  [Fig.~\ref{fig:4D-04}(c)].
\FloatBarrier
\subsubsection{Parametric exploration of the 4D Lorenz system's  dynamics}
\label{sec:4DODE_param}

\begin{figure}[htbp]\centering
\includegraphics[width=.45\columnwidth,keepaspectratio]{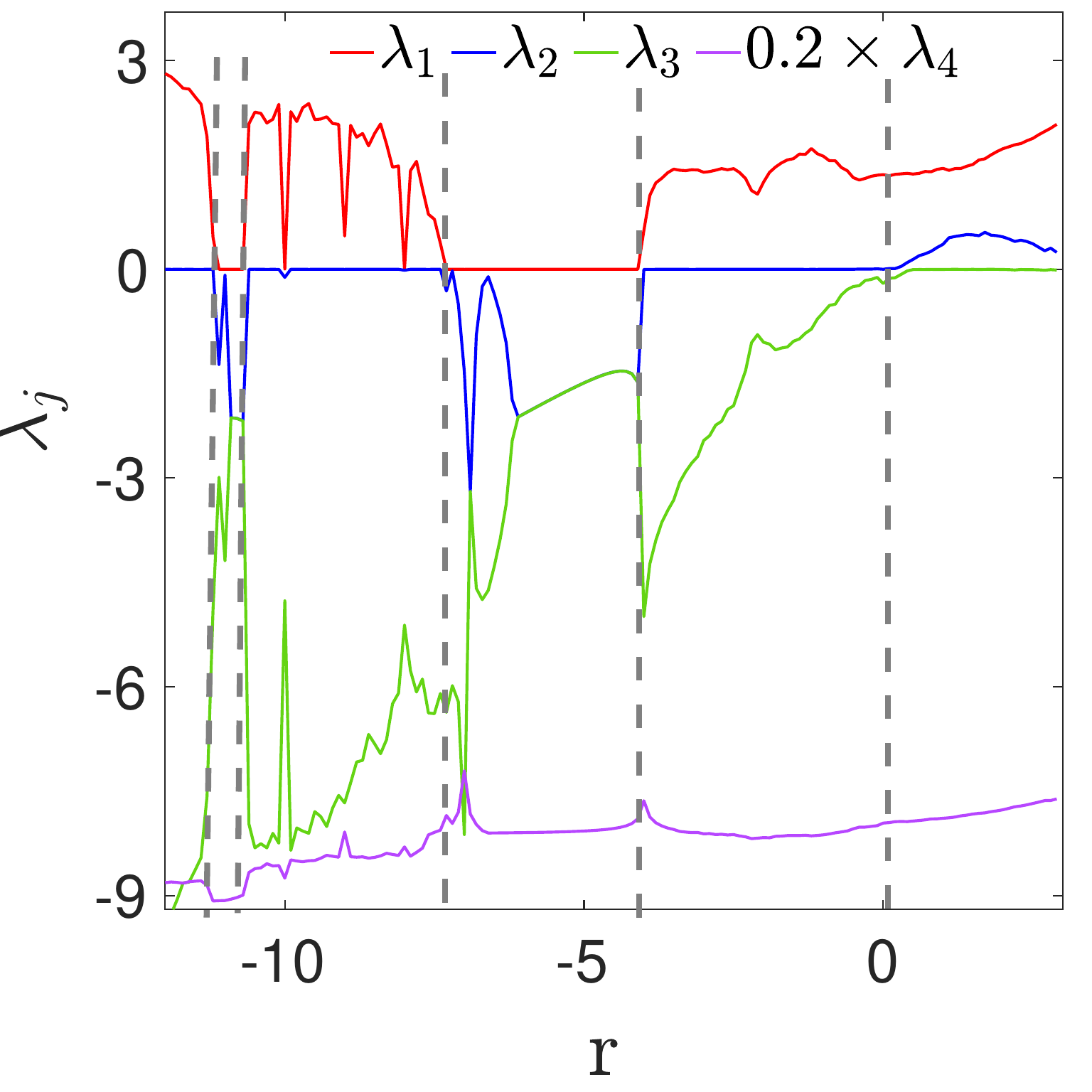}
\caption{The values of the ftLEs spectrum $\lambda_1$, $\lambda_2$, $\lambda_3$,  $\lambda_4$ (red, blue, green and purple curves respectively) at time $t=10^{4}$, as a function of the parameter $r$ of the 4D Lorenz system \eqref{eq:4DODE} with $a=35$, $b=8/3$, $c=55$, for the orbit with IC $(x,y,z,w) = (2,1,5,1)$.  The values of $\lambda_4$ have been rescaled for visualization purposes, while gray vertical dashed lines denote values $r=-11$, -10.65, -7.4, -4 and 0.1,  where transitions between different dynamical behaviors occur (see text for more details). }
\label{fig:4D-05}
\end{figure}

In order to study the dynamics of  the 4D Lorenz system \eqref{eq:4DODE} in a more extensive way,  we numerically investigate the fate of a representative trajectory with IC $(x,y,z,w) = (2,1,5,1)$ by setting $a=35$, $b=8/3$, $c=55$, and considering 300 equally spaced values of the parameter $r$ in the interval $[-12, 3]$. Integrating  this trajectory up to $t=10^{4}$  for each considered parameter set, and registering the value of its ftLEs after that time, we obtain the results presented in Fig.~\ref{fig:4D-05}. The computed values of the ftLEs ($\lambda_1$, $\lambda_2$, $\lambda_3$, and $\lambda_4$ are depicted by red, blue, green and purple curves respectively in Fig.~\ref{fig:4D-05}) allow us to identify parameter regions where different dynamical behaviors are observed. For $-12 \leq r \leq -11$ the motion is characterized by a positive $\lambda_1$, denoting the presence of chaotic motion, while for $-11 < r \leq -10.65$ $\lambda_1$ becomes approximately zero indicating the existence of stable limit cycles. Additionally,  chaotic attractors characterized by $\lambda_1>0$, $\lambda_2 \approx 0$ are observed for $r \in (-10.65,-7.4]$ and $r \in (-4,0.65]$, while stable limit cycles appear for $-7.4 < r \leq -4$. It is worth noting that for $0.1 < r \leq 3$ the two largest ftLEs,  $\lambda_1$ and $\lambda_2$, are positive, denoting the existence of  hyperchaotic motion. We also note that in Fig.~\ref{fig:4D-05} all the  above-mentioned $r$ values, where transitions between different dynamical behaviors occur, are denoted by vertical gray dashed lines. The results of Fig.~\ref{fig:4D-05} show that the computation of the spectrum of LEs allows the clear differentiation between diverse dynamical behaviors. On the other hand,  the GALI$_2$ index fails to identify these differences because it falls exponentially fast to zero, attaining very small values at the end of the integration time for all considered cases. This happens because for all trajectories considered in  Fig.~\ref{fig:4D-05},  $\lambda_1 > \lambda_2$, something which, according to \eqref{eq:GALI_chaos}, leads to the exponential decay of the GALI$_2$, as well of the GALIs of higher order.

\begin{figure}[h]\centering
\includegraphics[width=1.\columnwidth,keepaspectratio]{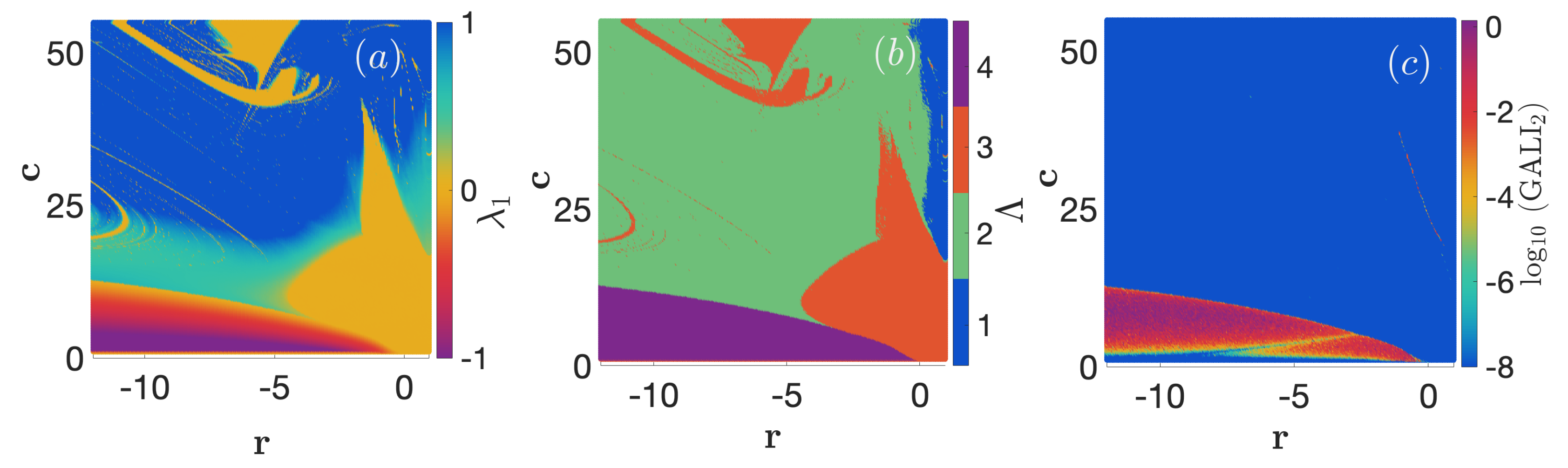}
\caption{The parameter space $(r,c)$  of the 4D Lorenz system \eqref{eq:4DODE} with $a=35$ and $b=8/3$,  colored according to the value of (a) the ftmLE $\lambda_1$ (scaled in the interval $[-1,1]$), (b) the index $\Lambda$, and (c) the GALI$_2$ of the trajectory with  IC $(x,y,z,w) = (3,2,10,1)$, at $t=10^{4}$. In (b)  the index  $\Lambda$ is  $\Lambda=1$ when $\lambda_1>0$, $\lambda_2 > 0$, $\lambda_3<0$, $\lambda_4<0$ (blue region), indicating the presence of hyperchaotic attractors, $\Lambda=2$ for $\lambda_1>0$, $\lambda_2 \leq 0$, $\lambda_3<0$, $\lambda_4<0$ (green region) corresponding to the appearance of chaotic attractors, $\Lambda=3$ when $\lambda_1 \approx 0$, $\lambda_2 < 0$, $\lambda_3<0$ $\lambda_4<0$  (orange region) denoting the existence of stable limit cycles, and $\Lambda=4$ when $\lambda_i <0$, $i=1,2,3,4$ (purple region) corresponding to the appearance of stable fixed points. Each color plot is created by considering  a set of $590 \times 260 = 153,400$ equally spaced grid points on the region  $(r, c) = [-12, 1] \times [1, 55]$.}
\label{fig:4D-Bipar}
\end{figure}

To perform an even broader investigation of system's \eqref{eq:4DODE} dynamical behavior we let two of its parameters  vary. Namely we consider setups for $r \in [-12, 1]$ and $c \in [1, 55]$, while $a$ and $b$ are kept fixed  to $a=35$ and $b=8/3$. For each one of these arrangements we follow the evolution of the trajectory with IC $(x,y,z,w) = (3,2,10,1)$ and  register the values of its ftLEs $\lambda_1$, $\lambda_2$, $\lambda_3$, and $\lambda_4$, as well as its GALI$_2$ at $t=10^{4}$. In Fig.~\ref{fig:4D-Bipar}(a) we color each point of the parameter space $(r,c)$ according to the trajectory's $\lambda_1$ value when it is scaled in the interval $[-1,1]$, as was also done in Fig.~\ref{fig:3D-Bipar}(a).   This process allows us to identify regions in the parameter space associated with the existence of different dynamical behaviors. More specifically, areas colored in yellow/orange ($\lambda_1 \approx 0$) indicate the presence of stable limit cycles, while purple/dark-red regions ($\lambda_1 < 0$)  denote the appearance of stable fixed points,  and  blue colored areas ($\lambda_1 > 0$) define parameter values for which chaotic or hyperchaotic attractors exist. 

The regions of the parameter space where different dynamical behaviors appear become more apparent if we use information about the whole spectrum of ftLEs and not only  $\lambda_1$,  as is done in Fig.~\ref{fig:4D-Bipar}(a).  Following a methodology similar to the one used for the creation of Fig.~\ref{fig:3D-Bipar}(b), we assign different values to an index $\Lambda$ as follows: $\Lambda=1$ when the final values of the ftLEs are arranged as $\lambda_1>0$, $\lambda_2 > 0$, $\lambda_3<0$, $\lambda_4<0$, denoting the presence of hyperchaotic attractors, $\Lambda=2$ for $\lambda_1>0$, $\lambda_2 \leq 0$, $\lambda_3<0$, $\lambda_4<0$,  corresponding to the appearance of chaotic attractors, $\Lambda=3$ when $\lambda_1 \approx 0$, $\lambda_2 < 0$, $\lambda_3<0$, $\lambda_4<0$, signifying the existence of stable limit cycles, and $\Lambda=4$ when  stable fixed point attractors  exist, and all  ftLEs are negative, i.e., $\lambda_i <0$, $i=1,2,3,4$. Based on this classification we color the system's parameter space  $(r,c)$ in Fig.~\ref{fig:4D-Bipar}(b) using 4 distinct colors for the different values of index $\Lambda$: 1 (blue), 2 (green), 3 (orange) and 4 (purple). 

From the results of Fig.~\ref{fig:4D-Bipar}(c),  where each point of the parameter space is colored according to the trajectory's GALI$_2$ value at $t=10^{4}$,   we see that the index is not able to differentiate between  regions of diverse dynamical behaviors,  as the majority of points are colored in blue, indicating very small values of the index (GALI$_2 \leq 10^{-8}$). This is due to the fact that, according to \eqref{eq:GALI_chaos}, whenever $\lambda_1> \lambda_2$, the GALI$_2$ exponentially decays to zero.  The only region of the parameter space where the GALI$_2$ does not become practically zero corresponds to the presence of stable fixed points, i.e., the region associated with $\Lambda=4$ colored in purple in Fig.~\ref{fig:4D-Bipar}(b), as in most cases there the two largest LEs, which are negative, are also practically equal.

\subsection{Numerical investigation of the generalized hyperchaotic H{\'e}non map} 
\label{sec:3DHenMap}

So far, in order to explore the GALIs' behavior in dissipative systems, we have considered examples of continuous time dynamical systems, namely the  3D \eqref{eq:3DODE} and 4D \eqref{eq:4DODE} Lorenz models. In all studied cases, we found that the time evolution of the GALIs is well described by \eqref{eq:GALI_chaos}. To further investigate the behavior of the GALIs for various types of trajectories occurring in  discrete time dissipative systems, we perform in this section a similar study to the ones presented in Secs.~\ref{sec:3DODE} and \ref{sec:4DODE}, for the generalized hyperchaotic H{\'e}non map \eqref{eq:3DHenMap}.

\subsubsection{A stable fixed point case}
\label{sec:Hen_fp}

\begin{figure}[htbp]\centering
\includegraphics[width=1\columnwidth,keepaspectratio]{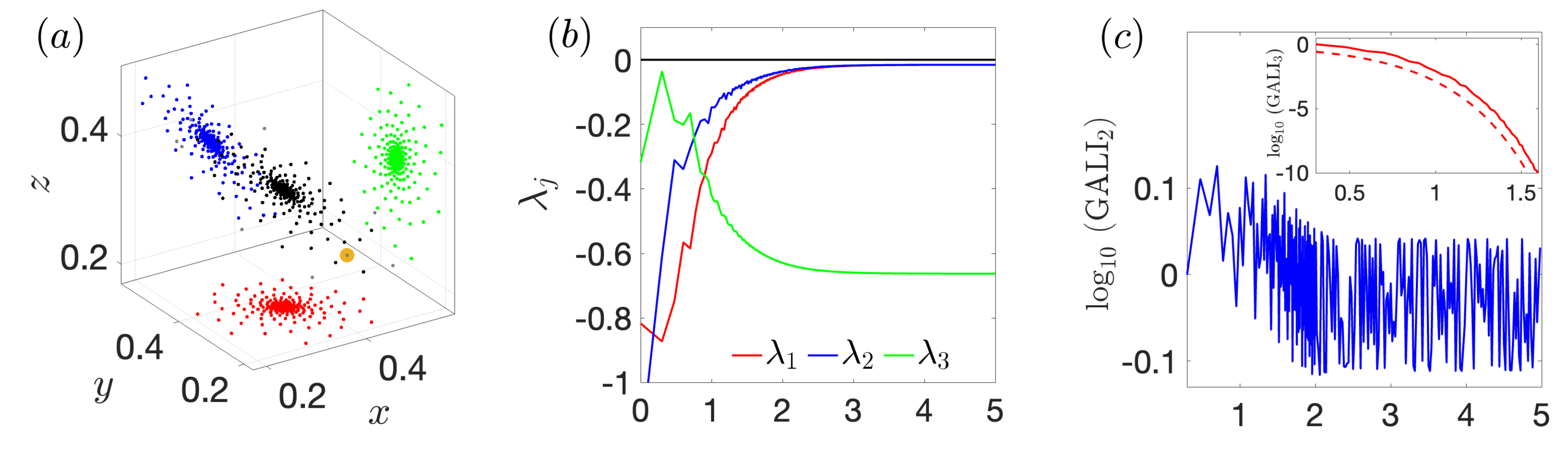}
\includegraphics[width=1\columnwidth,keepaspectratio]{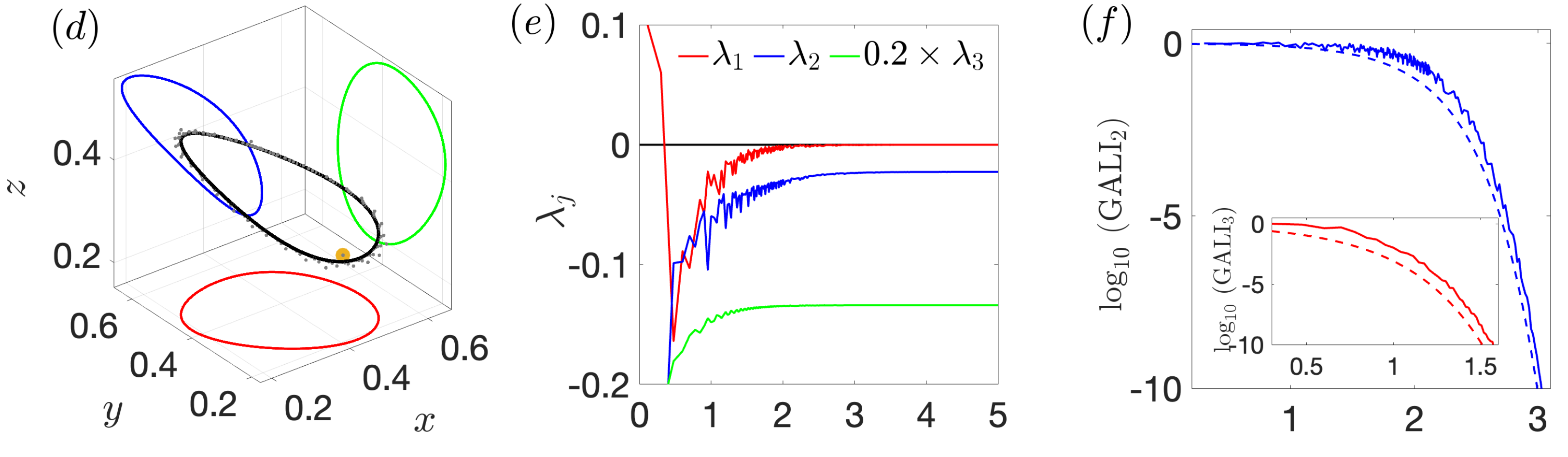}
\includegraphics[width=1\columnwidth,keepaspectratio]{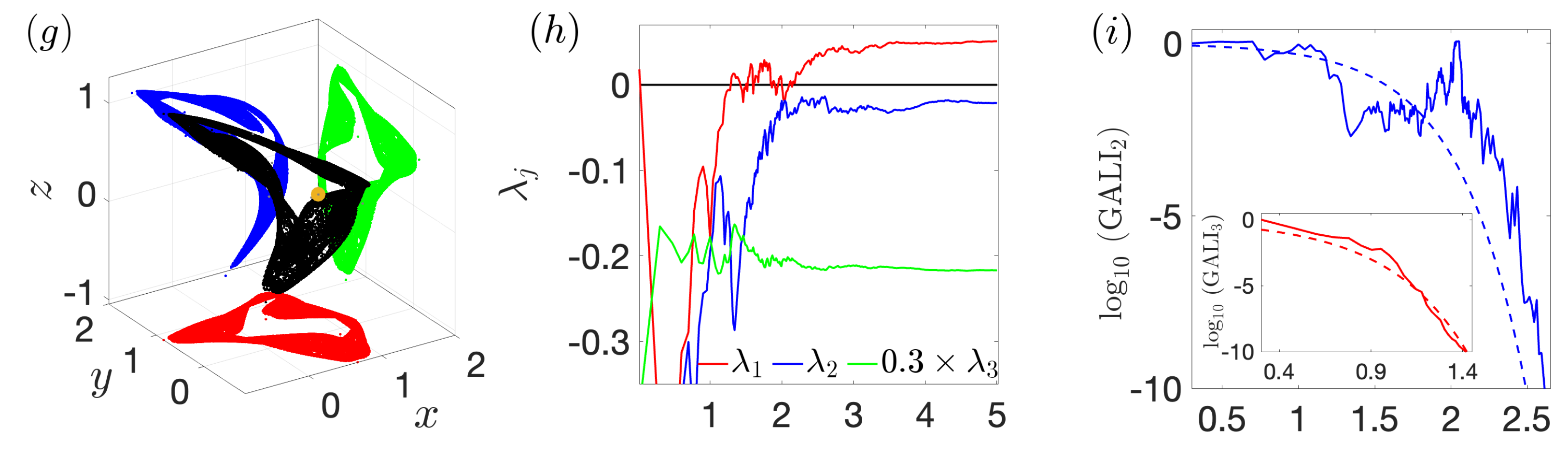}
\vspace{-0.5cm}
	\includegraphics[width=1\columnwidth,keepaspectratio]{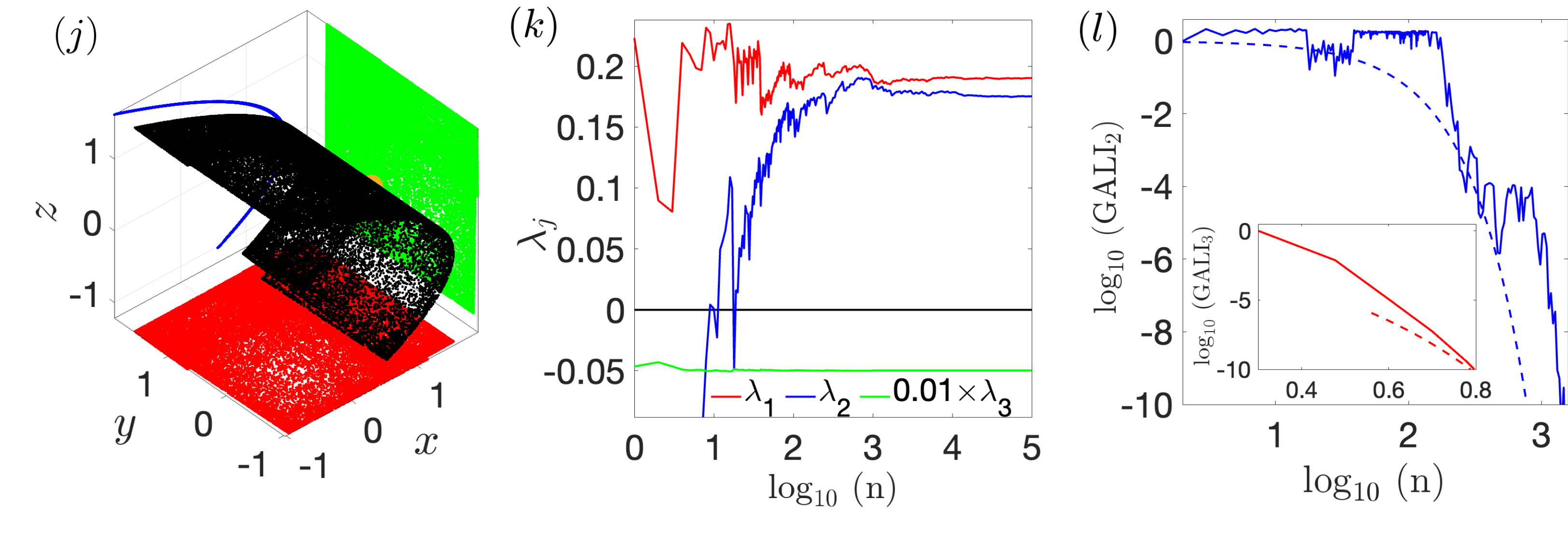}
	\caption{(Left column) 3D phase space portraits of trajectories with IC $(x, y, z) = (0.5, 0.4, 0.2)$ [indicated by an orange circle point in (a), (d), (g) and (j)],  for the generalized hyperchaotic H{\'e}non map \eqref{eq:3DHenMap} with parameters (a) $a=0.3$, $b=0.5$, (d) $a=0.3481$,  $b=0.5$, (g) $a=0.75$,  $b=0.01$, and (j) $a=1.6$, $b=0.01$. The trajectory asymptotically tends to (a) a stable fixed point, (d) a stable limit cycle,  (g) a chaotic  attractor, and (j) a hyperchaotic attractor. Gray points depict the initial part of the trajectory's evolution,  black points represent its asymptotic behavior, while  red,  blue and green points  show the orbit's  2D  $xy$, $xz$, and $yz$ projections  respectively. (Middle column) The time evolution of the ftLEs of the trajectories depicted in the respective panel of the left column: $\lambda_1$ (red curves), $\lambda_2$ (blue curves), and $\lambda_3$ (green curves). The black line in each panel indicates $\lambda_j=0$ for comparison. Note that in all panels  the $\lambda_3$ values have been scaled for visualization purposes. (Right column) The time evolution of the GALI$_2$ (solid blue  curves) and the GALI$_3$ (solid red  curves in the inset plots) for the orbits depicted in the first panel of each row.  Apart from the GALI$_2$ in (c),  which oscillates around a constant positive value, all GALIs decay exponentially fast to zero, following the functional forms (dashed curves) given in \eqref{eq:GALI_chaos}  for LEs' estimations obtained from the result presented in the middle column plots  (see text for the exact numerical values).  } 
	\label{fig:3DHM-01N}
\end{figure}

We begin our investigation by presenting in Fig.~\ref{fig:3DHM-01N}(a) the 3D phase space portrait of  the generalized hyperchaotic H{\'e}non map \eqref{eq:3DHenMap} with $a=0.3$ and $b=0.5$, for a trajectory with IC $(x, y, z) = (0.5, 0.4, 0.2)$ which approaches a stable fixed point attractor. From Fig.~\ref{fig:3DHM-01N}(a) we see that the trajectory's consequents (black points) tend to the stable fixed point $(x^*, y^*, z^*) = (0.3521, 0.3521, 0.3521)$  located at the center of the spiral created by the orbit's points. Fig.~\ref{fig:3DHM-01N}(b) shows that all the ftLEs of the trajectory are negative tending to the values $\chi_1=\chi_2=-0.015$ and $\chi_3=-0.66$. The fact that the two largest LEs  attain the same (negative) value results in the oscillations of the associated GALI$_2$ index around a positive value [blue curve in Fig.~\ref{fig:3DHM-01N}(c)]. On the other hand, the GALI$_3$ tends to zero exponentially fast [solid red  curve in the inset of Fig.~\ref{fig:3DHM-01N}(c)] following the exponential law GALI$_3 \propto \exp {\left[ -(2\chi_1 - \chi_2 - \chi_3) \right]}$ (dashed red curve) with $\chi_1=-0.015$, $\chi_2=-0.015$ and $\chi_3=-0.66$, in accordance with \eqref{eq:GALI_chaos}.

\subsubsection{A stable limit cycle case}
\label{sec:Hen_lc}

By setting $a=0.3481$ and $b=0.5$ in the  H{\'e}non map \eqref{eq:3DHenMap}, and keeping the IC to be $(x, y, z) = (0.5, 0.4, 0.2)$, we obtain a trajectory which tends to a stable limit cycle, as can be seen from the 3D phase space portrait of this orbit in Fig.~\ref{fig:3DHM-01N}(d). Figure \ref{fig:3DHM-01N}(e) shows the evolution of the trajectory's ftLEs. In this figure we see that  $\lambda_{1}$ (red curve) tends to zero,  while the other two ftLEs  (blue and green curves) remain always negative, approaching the values $\chi_2=-0.023$, and $\chi_3=-0.67$. Since $\chi_1 > \chi_2$, and in accordance with \eqref{eq:GALI_chaos}, both the GALI$_2$ and GALI$_3$ indices should exponentially tend  to zero. Indeed this is the case, as we can observe from the results of Fig.~\ref{fig:3DHM-01N}(f) where the time evolution of the GALI$_2$ (solid blue curve) and the GALI$_3$ (solid red curve in the inset of the figure) is shown. These exponential decays are very well approximated by $\exp {\left[ -(\chi_1 - \chi_2) \right]}$ (dashed blue curve) and  $\exp {\left[ -(2\chi_1 - \chi_2 - \chi_3) \right]}$ (dashed red curve in the figure's inset) for $\chi_1 = 0$, $\chi_2=-0.023$, and $\chi_3=-0.67$.

\subsubsection{A chaotic, strange attractor case}
\label{sec:Hen_sa}

By changing the parameters of the  H{\'e}non map \eqref{eq:3DHenMap} to $a=0.75$ and $b=0.01$, the trajectory with IC  $(x, y, z) = (0.5, 0.4, 0.2)$ yields to a chaotic attractor [Fig.~\ref{fig:3DHM-01N}(g)], characterized by an eventually positive ftmLE $\lambda_1 =0.051$, while $\lambda_{2}$ and $\lambda_{3}$ remain negative, asymptotically attaining the values $\chi_2 = -0.021$, and $\chi_3 = -0.72$ respectively [Fig.~\ref{fig:3DHM-01N}(h)].  Since $\chi_1 > \chi_2$, similarly to what was observed for the trajectory of Sec.~\ref{sec:Hen_lc},  both the GALI$_2$ [solid blue curve in Fig.~\ref{fig:3DHM-01N}(i)] and the GALI$_3$ [solid red curve in the inset of Fig.~\ref{fig:3DHM-01N}(i)] decrease to zero, following exponential decay rates defined in \eqref{eq:GALI_chaos}, namely GALI$_2 \propto \exp {\left[ -(\chi_1 - \chi_2) \right]}$ and GALI$_3 \propto \exp {\left[ -(2\chi_1 - \chi_2 - \chi_3) \right]}$ [blue and red dashed curves in Fig.~\ref{fig:3DHM-01N}(i) and its inset respectively] for $\chi_1 = 0$, $\chi_2 = -0.021$, and $\chi_3 = -0.72$.

\subsubsection{A hyperchaotic attractor case}
\label{sec:Hen_hyper}
 
Figure \ref{fig:3DHM-01N}(j) illustrates the phase space portrait of the trajectory with IC $(x, y, z) = (0.5, 0.4, 0.2)$  of the H{\'e}non map \eqref{eq:3DHenMap} with $a=1.6$ and $b=0.01$.  This trajectory tends to a hyperchaotic attractor, similarly to what was also observed in Fig.~3 of  \cite{wang2023image}, but for different parameter values of \eqref{eq:3DHenMap}. The hyperchaotic nature of the attractor is reflected on the fact that, as is seen in Fig.~\ref{fig:3DHM-01N}(k), the trajectory has two positive ftLEs tending to values $\chi_1 = 0.19$ and $\chi_2 = 0.18$, with the third one being negative, tending to $\chi_3 = -4.97$. Again, due to the fact that $\chi_1 > \chi_2$ the GALI$_{2}$ and the GALI$_{3}$ decrease to zero exponentially fast, i.e., 
GALI$_2 \propto \exp {\left[ -(\chi_1 - \chi_2) \right]}$ and GALI$_3 \propto \exp {\left[ -(2\chi_1 - \chi_2 - \chi_3) \right]}$ [Fig.~\ref{fig:3DHM-01N}(l)].

\subsubsection{Parametric exploration of the dynamics of the generalized hyperchaotic H{\'e}non map}
\label{sec:Hen_param}

\begin{figure}[htbp]\centering
	\includegraphics[width=1.\columnwidth,keepaspectratio]{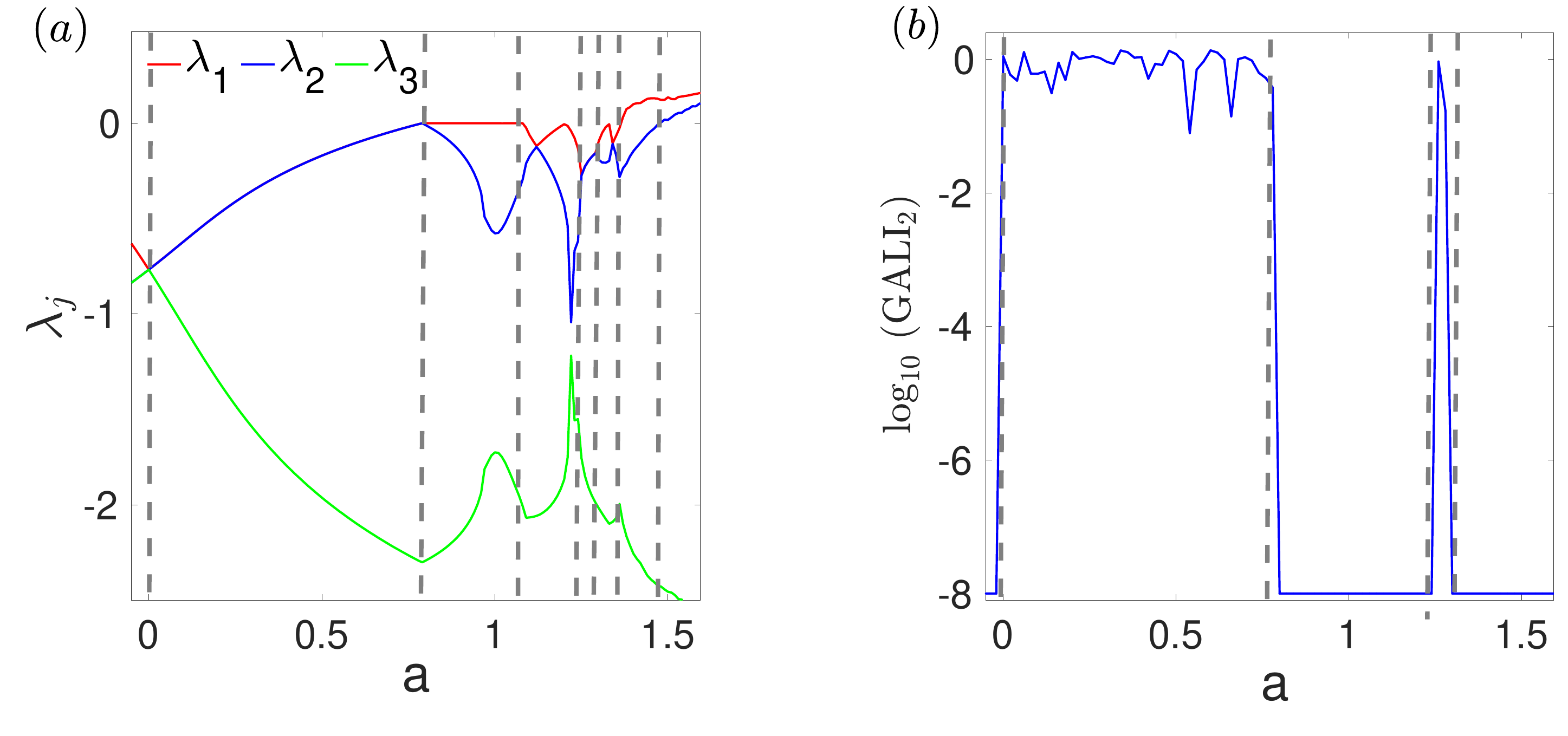}
    \caption{The values, after $n=10^4$  iterations  of the H{\'e}non map \eqref{eq:3DHenMap} with $b=0.1$  of (a) the spectrum of the ftLEs $\lambda_1$,  $\lambda_2$, $\lambda_3$ (red, blue, and green curves respectively),  and (b) the GALI$_2$,   as a function of $a$ ($a \in [-0.05, 1.6]$)  for the trajectory with IC $(x, y, z) = (0.5, 0.4, 0.2)$.  Gray vertical dashed lines denote in (a) the values $a=0.01$, $0.7835$, $1.0835$, $0.125$, $0.129$, $1.3634$ and $1.4835$ in (a), and  the values $a=0.01$, $0.7835$, $0.125$, $0.129$ in (b). }
    \label{fig:3DHM-05}
\end{figure}

\begin{figure}[htbp]\centering
	\includegraphics[width=1.\columnwidth,keepaspectratio]{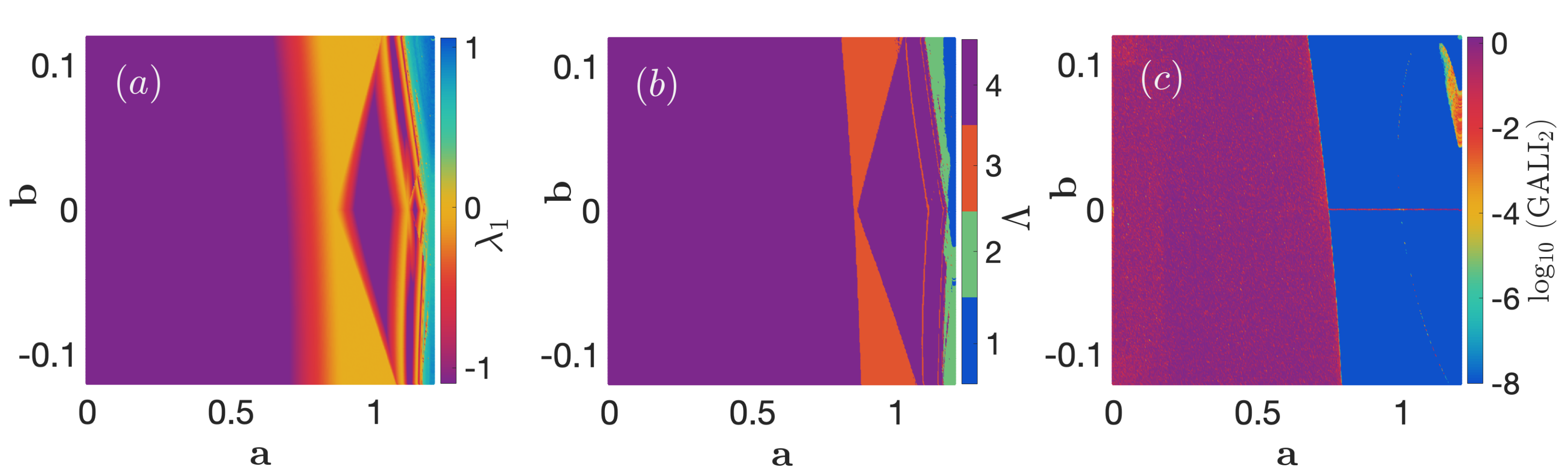}
	\caption{The parameter space $(a, b)$  of the  generalized hyperchaotic H{\'e}non map \eqref{eq:3DHenMap},  colored according to the value of (a) the ftmLE $\lambda_1$ (scaled in the interval $[-1,1]$), (b) the index $\Lambda$, and (c) the GALI$_2$ of the trajectory with IC $(x, y, z) = (0.5, 0.4, 0.2 )$, after $n=10^4$ iterations. In (b)  the index  $\Lambda$ is  $\Lambda=1$ when $\lambda_1 >0$, $\lambda_2 >0$, $\lambda_3 <0$ (blue region), indicating the presence of hyperchaotic attractors, $\Lambda=2$ for $\lambda_1 >0$, $\lambda_2 \leq 0$, $\lambda_3 <0$ (green region) corresponding to the appearance of chaotic attractors, $\Lambda=3$ when $\lambda_1 \approx 0$, $\lambda_2 < 0$, $\lambda_3 <0$ (orange region) denoting the existence of stable limit cycles, and $\Lambda=4$ when $\lambda_i <0$, $i=1,2,3$ (purple region) corresponding to the appearance of stable fixed points. Each color plot is created by considering  a set of $240 \times 1,765 = 423,600$ equally spaced grid points on the region  $(a, b) = [0, 1.2] \times [-0.12, 0.12]$. }
	\label{fig:3DHM-Bipar}
\end{figure}

Similarly to what was done in Figs.~\ref{fig:3D-02} and \ref{fig:3D-Bipar}  for the 3D Lorenz system   \eqref{eq:3DODE}, and in Figs.~\ref{fig:4D-05} and \ref{fig:4D-Bipar}  for the 4D Lorenz system   \eqref{eq:4DODE}, we now perform a more global analysis of the dynamics of the H{\'e}non map \eqref{eq:3DHenMap}, by varying only one [Fig.~\ref{fig:3DHM-05}], or both its parameters [Fig.~\ref{fig:3DHM-Bipar}].

The results of Fig.~\ref{fig:3DHM-05} are obtained by considering the trajectory with IC $(x, y, z) = (0.5, 0.4, 0.2 )$,  fixing $b=0.1$,  and varying the values of $a$ in the interval $[-0.05, 1.6]$.  More specifically, in Fig.~\ref{fig:3DHM-05}(a) we present the values of the trajectory's ftLEs, $\lambda_1$, $\lambda_2$ and $\lambda_3$ (red, blue and green curves respectively) after $n=10^4$ iterations of the map,  as a  function of $a$, while in Fig.~\ref{fig:3DHM-05}(b) we have a similar plot for the values of the GALI$_2$. In Fig.~\ref{fig:3DHM-05}(a) we see that for $a \leq  0.7835$ the trajectory tends to a stable fixed point, like the one depicted in Fig.~\ref{fig:3DHM-01N}(a), and consequently all its ftLEs are negative.  For $ 0.7835 < a \leq 1.0835$ we practically have $\lambda_1 = 0$, indicating the presence of a stable limit cycle,  similar to the one shown  in Fig.~\ref{fig:3DHM-01N}(d). Then, for $1.0835 < a \leq 1.3634$  the system exhibits again fixed points characterized by all ftLEs  being negative. For $a > 1.3634$ the ftmLE, $\lambda_1$, becomes positive (while both $\lambda_2$ and $\lambda_3$ are negative),  denoting the presence of a chaotic attractor, while for $a > 1.4835$ hyperbolic behavior appears as the two largest ftLEs are positive. As can be seen in Fig.~\ref{fig:3DHM-05}(b) the GALI$_2$ is different from zero only when the  two largest ftLEs are practically equal. This happens in the ranges $0.01 \leq a \leq  0.7835$  and $0.125 \leq a \leq  0.129$,  where stable fixed point attractors exist. It is worth noting that for $ a \in [-0.05, 0.01]$, where again stable fixed points appear, as it can be understood from the negative values  of all ftLEs in Fig.~\ref{fig:3DHM-05}(a),  the GALI$_2$ becomes again  zero, in accordance with \eqref{eq:GALI_chaos}, because $\lambda_1 \neq \lambda_2$.
\FloatBarrier

In Fig.~\ref{fig:3DHM-Bipar} we present the results obtained in the parameter space of the the H{\'e}non map \eqref{eq:3DHenMap},   defined by $a \in [0, 1.2]$ and $b \in [-0.12, 0.12]$,  when a grid consisting of 240 and 1,765 equally spaced points  along the $a$ and $b$ axis respectively is considered. For each parameter set the orbit with IC $(x, y, z) = (0.5, 0.4, 0.2 )$  is iterated $n=10^4$  times and its set of ftLEs and GALI$_2$ values are computed.  In Fig.~\ref{fig:3DHM-Bipar}(a) we color points according to the related ftmLE, $\lambda_1$, value scaled in the range $[-1,1]$, as was also done in Figs.~\ref{fig:3D-Bipar}(a) and \ref{fig:4D-Bipar}(a).  Purple colored regions ($\lambda_1 <0$)  denote parameter sets leading to stable fixed points, yellow/orange areas corresponding to $\lambda_1 \approx0$ indicate the existence of stable limit cycles, and blue  areas correspond to the presence of chaotic motion. A clearer distinction between the regions of the parameter space where different dynamical behaviors occur is achieved in Fig.~\ref{fig:3DHM-Bipar}(b) where points are colored according to the value of the $\Lambda$ index, which depends on the arrangement of the whole spectrum of ftLEs. In particular, blue regions ($\Lambda=1$) denote the existence of hyperbolic motion ($\lambda_1 >0$, $\lambda_2 >0$, $\lambda_3 <0$), green areas ($\Lambda=2$) signify chaotic behavior ($\lambda_1 >0$, $\lambda_2 \leq 0$, $\lambda_3 <0$), orange  colored regions ($\Lambda=3$) correspond to stable limit cycles ($\lambda_1 \approx 0$, $\lambda_2 < 0$, $\lambda_3 <0$), and  purple regions ($\Lambda=4$) indicate the presence of stable fixed point attractors ($\lambda_i <0$, $i=1,2,3$). As was also observed in Figs.~\ref{fig:3D-Bipar}(c) and \ref{fig:4D-Bipar}(c), the GALI$_2$ fails to clearly  discriminate between parameter regions where different attractors appear, as it attains very small values [in practice the index becomes zero -  blue regions Fig.~\ref{fig:3DHM-Bipar}(c)] in all cases for which $\lambda_1 \neq \lambda_2$. Whenever the two largest ftLEs are practically equal the index oscillates around positive values [purple colored areas in Fig.~\ref{fig:3DHM-Bipar}(c)]. We note that some parameter sets in the upper right corner of Fig.~\ref{fig:3DHM-Bipar}(c) are  colored in yellow/red/green, indicating the existence of weakly chaotic trajectories, which  require more iterations for the GALI$_2$ to decay to zero and to clearly reveal the motion's chaotic nature. 

\section{Summary and discussion} 
\label{sec:Summary}

In this work we investigated in detail the behavior of the GALI method for different, typical types of motion encountered in dissipative dynamical systems. By doing that, we completed, in some sense, the study of the GALI technique across the spectrum of dynamical systems, since the method has already extensively, and very successfully, been used as chaos indicator in conservative Hamiltonian systems and area preserving maps. 

In our investigation we considered two continuous time dissipative dynamical models, namely the 3D \eqref{eq:3DODE} and the 4D \eqref{eq:4DODE} Lorenz systems, as well as the generalized hyperchaotic H{\'e}non map \eqref{eq:3DHenMap}, which is a discrete time model. Using the computation of the mLE, as well as of the whole spectrum of LEs, we identified individual trajectories of diverse dynamical behaviors, i.e, orbits leading to stable fixed points, stable limit cycles, chaotic and hyperchaotic attractors. Furthermore, we also defined   regions in the parameter spaces of the studied models where these different types of attractors exist. Our studies showed that the computation of the whole spectrum of LEs, or even the estimation of only the mLE (something which is obviously computationally less demanding)  manages to correctly discriminate between the different types of motions.

With respect to the performance of the GALI method we found that the time evolution of the index is always well approximated by \eqref{eq:GALI_chaos}, which indicates an exponential decay of its values. We stress that \eqref{eq:GALI_chaos} dictates that the GALI of order $k$ (GALI$_k$) will remain practically constant [or in other words, will follow the behavior described in \eqref{eq:GALI_regular}] if the first $k$ LEs are equal. In our extensive numerical simulations we found cases where this is true only for $k=2$. Thus, the GALI$_k$ with $k>2$ decreased exponentially fast to zero for all considered types of motions. This behavior clearly indicates that the GALI$_k$ with $k>2$ cannot be used to discriminate between different types of trajectories in dissipative dynamical systems.

On the other hand, the GALI$_2$ did exhibit diverse behaviors. The index fluctuated around a practically constant positive value for orbits tending to stable fixed point attractors, as these trajectories were typically characterized by having their two largest LEs, $\chi_1$ and $\chi_2$, attaining negative, but nevertheless practically equal, values. Furthermore, the GALI$_2$ decayed exponentially fast to zero for trajectories tending to limit cycles, as well as chaotic and hyperchaotic attractors, because the two largest LEs of these orbits were not equal ($\chi_1 > \chi_2$). Consequently, the GALI$_2$ cannot discriminate between these dynamical behaviors. Thus, it is advisable for the GALI$_2$ to be used with caution for studies or dissipative dynamical systems, and  preferably in conjunction with the computation of the mLE, or even the whole spectrum of LEs.

\section*{Acknowledgments} 

\noindent H.~T.~M.~acknowledges funding from the Science Faculty PhD Fellowship of the University of Cape Town (UCT), the Ethiopian Ministry of Education and Woldia University, as well as partial support from the UCT Incoming International Student Award and the Science Faculty Postgraduate Publication Incentive (PPI) funding. We thank the High Performance Computing facility of the University of Cape Town and the Centre for High Performance Computing (CHPC) of South Africa for providing computational resources for this project.


\bibliographystyle{plain}  
\bibliography{MMRS}

\end{document}